\documentclass[prb,aps,twocolumn,showpacs]{revtex4}

\usepackage{graphicx,epsfig}
\usepackage{bm}
\usepackage{multirow}
\usepackage{amsmath}

\begin{document}

\title{Crossover from adiabatic to antiadiabatic
       phonon-assisted tunneling in single-molecule
       transistors}
\smallskip

\author{Eitan Eidelstein, Dotan Goberman, and
        Avraham Schiller}

\affiliation{Racah Institute of Physics, The Hebrew
             University, Jerusalem 91904, Israel}

\begin{abstract}
The crossover between two customary limits of
phonon-assisted tunneling, the adiabatic and
antiadiabatic regimes, is studied systematically in
the framework of a minimal model for molecular devices:
a resonant level coupled by displacement to a localized
vibrational mode. Conventionally associated with the
limits where the phonon frequency is either sufficiently
small or sufficiently large as compared to the bare
electronic hopping rate, we show that the crossover
between the two regimes is governed for
strong electron-phonon interactions primarily
by the polaronic shift rather than the phonon
frequency. In particular, the perturbative adiabatic
limit is approached only as the bare hopping rate
$\Gamma$ exceeds the polaronic shift, leaving an
extended window of couplings where $\Gamma$ well
exceeds the phonon frequency and yet the physics
is basically that of the antiadiabatic regime.
We term this intermediate regime the extended
antiadiabatic regime. The effective low-energy
Hamiltonian in the traditional and the extended
antiadiabatic regime is shown to be the (purely
fermionic) interacting resonant-level model,
with parameters that we extract from numerical
renormalization-group calculations. The extended
antiadiabatic regime is followed in turn by a
true crossover region where the polaron gets
progressively undressed. In this latter region, the
phonon configuration strongly deviates from a simple
superposition of just one or two coherent states.
The renormalized tunneling rate, which serves
as the low-energy scale in the problem and thus
sets the width of the tunneling resonance, is
found to follow an approximate scaling form on
going from the adiabatic to the antiadiabatic regime.
Charging properties are governed by two distinct
mechanisms at the extended antiadiabatic and into
the crossover region, giving rise to characteristic
shoulders in the low-temperature conductance as
a function of gate voltage. These shoulders serve
as a distinct experimental fingerprint of
phonon-assisted tunneling when the electron-phonon
coupling is strong.
\end{abstract}



\maketitle

\section{Introduction}
\label{Sec:Introduction}

The promise of molecular electronics has
focused enormous interest on molecular
devices.~\cite{Reviews} Typically, such devices consist
of an individual molecule trapped between two leads,
in-between which a voltage bias is applied. By measuring
the current flowing across the molecular bridge one
can investigate the molecule's internal degrees of
freedom and their coupling to the leads, which can
lead in turn to complex many-body physics.
While lacking the exquisite design and control
capabilities of semiconductor quantum dots, molecular
devices can be produced in large quantities, thus
allowing for many samples to be scanned in a relatively
short period of time. At the same time, it is not
always clear if a molecule has been successfully
trapped between the leads, or whether it might have
been damaged or distorted in the course of preparation.

From a basic-science perspective, single-molecule
transistors (SMT) offer two major advantages over
their semiconductor counterparts. First, the relevant
energy scales are notably larger in SMTs, rendering
these scales more accessible to experiments. Second, the
electronic degrees of freedom are generally coupled to
nuclear vibrational modes, providing an extraordinary
opportunity to study the electron-phonon coupling at
the nano-scale. The same picture applies to suspended
carbon nanotubes, where the motion of electrons is
coupled to vibrations of the tube (vibrons). Indeed,
phonon-assisted tunneling can lead to a plethora of
interesting phenomena, including the appearance of
inelastic steps and peaks in the differential
conductance,~\cite{ZMB02,QNH04,Natelson-04,Sapmaz-etal-06,
Tal-etal-08} the Frank-Condon
blockade,~\cite{Sapmaz-etal-06,Leturcq-etal-09}
and the interplay with the Kondo
effect.~\cite{YN04,Parks-et-al-07,Natelson-review}
While most of the experiments cited above are in
general accord with theoretical expectations, some
issues, such as the sign of the inelastic steps at
integer multiples of the phonon frequency, remain
under debate.~\cite{EG08,EWIA09} Other theoretical
predictions awaiting experimental verification
include unorthodox variants of the Kondo
effect,~\cite{Cornaglia-et-al,Two-channel-SMT}
signatures of pair tunneling,~\cite{KRvO06} and
interesting nonequilibrium effects on the phonon
distribution function,~\cite{Koch-et-al-06} to
name but a few.

While much of the theoretical activity on molecular
devices is presently centered on finite-bias transport,
in this paper we focus on thermal equilibrium and
address a specific question pertaining to the nature
of the crossover between two customary limits of
phonon-assisted tunneling, the so-called adiabatic
and antiadiabatic regimes. These two terms are
broadly used in the context of electron-phonon
coupling to indicate the limits where the bare
electronic motion is either sufficiently fast
(adiabatic limit) or sufficiently slow
(antiadiabatic limit) as compared to the phonon
vibrations. In the more specific context of
resonant phonon-assisted tunneling, the relevant
measure of electronic motion is given by the
tunneling rate. Hence, the two limits correspond
to whether the bare electronic tunneling rate
$\Gamma$ is either sufficiently small or sufficiently
large as compared to the phonon frequency $\omega_0$.

When $\Gamma \ll \omega_0$ (we work with units in
which $\hbar = 1$), the phonon can efficiently
respond to hopping events by forming a polaron,
suppressing thereby the electronic tunneling rate.
This suppression, which can be quite dramatic, is
manifest, e.g., in a narrowing of the tunneling
resonance. In the opposite limit, $\omega_0 \ll \Gamma$,
the phonon is too slow to respond to the frequent
tunneling events, having little effect on their
rate. Each of these extreme limits is rather well
controlled theoretically, either in the framework of
the Lang-Firsov transformation~\cite{LF62} or using
ordinary perturbation theory in the electron-phonon
coupling. Far less understood is the crossover
region between the two limits, which lacks a
small parameter.

This general picture neglects, however, another
important energy scale: the harmonic potential
energy associated with the relative displacement of
the phonon between different molecular electronic
configurations. In the case of a single spinless level
with the dimensionless displacement coupling $\lambda$
[see Eq.~(\ref{H-ep}) for an explicit definition of
$\lambda$], the so-called polaronic shift is given
by $E_{\rm p} = \lambda^2 \omega_0$. Thus,
depending on the magnitude of $\lambda$, the
polaronic shift $E_{\rm p}$ may exceed
$\Gamma$ and/or $\omega_0$. Quantum mechanically it
is natural to associate $E_{\rm p}$ with a new time
scale $\tau_{\rm p} = 1/E_{\rm p}$, which may
either be shorter or longer than the electronic dwell
time $\tau_{\rm dwell} = 1/\Gamma$ and the period
of oscillations $\tau_{\rm osc} = 2 \pi/\omega_0$.
Whether $\tau_{\rm p}$ has the status of a true
physical time scale is not immediately clear. It
can not have real significance for $\lambda \ll 1$,
when the phonon displacement is small as compared to
its zero-point motion. Neither does $\tau_{\rm p}$
play any role for a classical oscillator, whose period
is independent of the amplitude of oscillations. At
the same time, $E_{\rm p}$ does show up as
an additional energy scale for phonon-assisted
tunneling,~\cite{Selman} although its significance, let
alone its role in defining the physical boundaries
of the adiabatic and antiadiabatic regimes, have
never been quite resolved.

From this brief discussion it is clear that the most
interesting and yet most challenging regime is that
of strong electron-phonon coupling, $1 \ll \lambda$,
where $\tau_{\rm p}$, whatever its physical
interpretation might be, can potentially assume a
dominant role. The interest in strong electron-phonon
coupling is further amplified by recent reports of
large values of $\lambda$ (including some in excess of
5) in suspended carbon nanotubes.~\cite{Leturcq-etal-09}
In the antiadiabatic limit, the tunneling
electrons experience strong polaronic dressing for
$1 \ll \lambda$, reflected in an exponential
suppression of the renormalized tunneling rate from
$\Gamma$ to $\Gamma_{\rm eff} = \Gamma e^{-\lambda^2}
\ll \Gamma$. This dramatic effect raises several
basic questions:
\begin{enumerate}
\item
    When tuning the bare tunneling rate from weak
    ($\Gamma \ll \omega_0$) to strong
    ($\omega_0 \ll \Gamma$) coupling, which physical
    parameters set the scale for first leaving the
    polaronic physics of the antiadiabatic regime and
    then entering the perturbative physics of the
    adiabatic regime? Are these two transitions
    governed by a single scale or are there perhaps
    two distinct scales?
\item
     How does the ratio $\Gamma_{\rm eff}/\Gamma$
     evolve from its exponentially small value
     $e^{-\lambda^2}$ in the antiadiabatic regime
     back to approximately one in the adiabatic
     limit? In other terms, how does the polaron
     get undressed?
\item
     Are there any distinct experimental signatures
     of the crossover regime that can be detected?
\end{enumerate}

\subsection{Preliminaries}

In this paper, we answer these questions in detail
in the framework of the resonant-level model with
an additional displacement coupling to a localized
vibrational mode [see Eq.~(\ref{H-ep})].
Besides being one of the most widely used models
for single-molecule devices, our motivation for
adopting this specific Hamiltonian is two-fold.
The first is physical clarity, as this Hamiltonian
constitutes the minimal model where the crossover
from the antiadiabatic to the adiabatic regime of
phonon-assisted tunneling can be studied without
being masked by competing many-body effects (e.g.,
the Kondo effect in case of a spinful level). The
second point is technical in nature, and pertains to
our method of choice for accurate nonperturbative
calculations. In this work we employ Wilson's
numerical renormalization-group (NRG)
approach,~\cite{Wilson75,BCP08} which is a highly
precise tool for calculating equilibrium properties
of quantum impurity systems. In the NRG, the
computational effort grows exponentially with the
number of conduction-electron species, hence the
restriction to a single spinless band allows us to
accurately address large values of $\lambda$ that
otherwise would be inaccessible using more elaborate
models.

\begin{figure}[t]
\centerline{
\includegraphics[width=85mm]{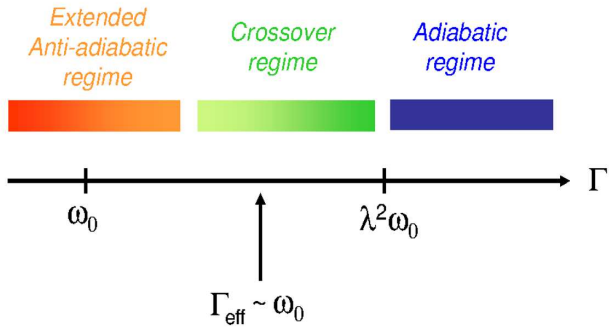}
}\vspace{0pt}
\caption{(Color online)
         Schematic description of the crossover from
         the antiadiabatic to the adiabatic regime
         of resonant phonon-assisted tunneling, for
         strong electron-phonon interactions
         $1 \ll \lambda$. Upon increasing $\Gamma$,
         an extended antiadiabatic regime persists
         until the renormalized tunneling rate
         $\Gamma_{\rm eff}$ approaches the phonon
         frequency $\omega_0$. The condition
         $\Gamma_{\rm eff} = \omega_0$ is typically
         met for $\Gamma \sim 0.55 E_{\rm p}$ with
         $E_{\rm p} = \lambda^2 \omega_0$, extending
         the physics of the antiadiabatic regime
         far beyond the traditional condition
         $\Gamma < \omega_0$. In terms of the bare
         model parameters, the extended antiadiabatic
         regime persists up to $\Gamma \sim 0.4 E_{\rm p}$.
         The perturbative physics of the adiabatic regime is
         approached only as $\Gamma$ exceeds $E_{\rm p}$,
         leaving an intermediate crossover region
         for $0.4 E_{\rm p} \alt \Gamma \alt E_{\rm p}$
         where the polaron gets progressively undressed.
         The phonon configuration strongly deviates in
         the crossover region from a simple superposition
         of just one or two coherent states.}
\label{Fig:E-p}
\end{figure}

We note in passing that the NRG has been applied to
moderately large values of $\lambda$ in the framework
of the spinful Anderson-Holstein model,~\cite{HM02}
however, these studies used rather large values of
the phonon frequency. The combination of large
values of $\lambda$ and small phonon frequencies (as
is appropriate for molecular and nanotube devices) is
presently a challenge to treat accurately using the
NRG if spin is to be included.

Focusing on $1 \ll \lambda$, Fig.~\ref{Fig:E-p}
displays our resulting scenario for the crossover
from the antiadiabatic to the adiabatic regime.
In contrast to the naive picture, the crossover
is governed primarily by the polaronic shift
$E_{\rm p}$ rather than the phonon frequency
$\omega_0$. In particular, the perturbative adiabatic
limit is approached only as $\Gamma$ exceeds
$E_{\rm p}$, leaving a rather broad window of
couplings where $\Gamma$ well exceeds $\omega_0$
and yet the physics is basically that of the
antiadiabatic regime. We term this intermediate
regime the {\em extended antiadiabatic} regime.

The physical origin of the extended antiadiabatic
regime is in the enormous separation of scales
between $\Gamma$ and $\Gamma^{}_{\rm eff}$ when
$\Gamma \ll E_{\rm p}$. Although smaller than
the bare tunneling rate, the phonon frequency
$\omega_0$ remains notably larger than the
renormalized one, hence the phonon can still
efficiently respond to the individual tunneling
events. The distinction between the extended and
the traditional antiadiabatic regimes is therefore
quantitative rather than qualitative, both being
described by the same fermionic interacting
resonant-level model at energies below $\omega_0$.
For this reason, we have lumped the two regimes
into one in Fig.~\ref{Fig:E-p}.
In terms of the bare model parameters, the
extended antiadiabatic regime persists up to
$\Gamma \sim 0.4 E_{\rm p}$, breaking down as
$\Gamma^{}_{\rm eff}$ approaches $\omega_0$ in
magnitude. The extended antiadiabatic regime
and the adiabatic one are separated in turn
by a true crossover region
for $0.4 E_{\rm p} \alt \Gamma \alt E_{\rm p}$,
where the polaron gets progressively undressed.
The phonon configuration strongly deviates in this
region from a simple superposition of just one or
two coherent states, as we show by explicit
calculations.

\subsection{Plan of the paper}

After introducing the Hamiltonian and its
symmetries in Sec.~\ref{Sec:model}, we proceed in
Sec.~\ref{Sec:limits} to a preliminary discussion
of certain limits where analytic insight can be
gained. These include the traditional adiabatic and
antiadiabatic limits, the extended antiadiabatic
regime proposed in this paper, and the limit of
large detuning. Section~\ref{Sec:NRG-study} presents
in turn a systematic study of all coupling regimes
using Wilson's NRG. To this end, we begin with a
brief introduction of the method in Sec.~\ref{Sec:NRG},
followed by an extensive investigation of the key
quantities of interest: the renormalized tunneling
rate $\Gamma_{\rm eff}$, the mapping onto an
effective low-energy interacting resonant-level
model in the antiadiabatic and the extended
antiadiabatic regimes, charging of the level,
and the phonon distribution function defined in
Eq.~(\ref{P(n)-def}). We conclude in
Sec.~\ref{Sec:summary} with a summary of our
results. Some technical details are deferred
to two appendices.

\section{The model and its symmetries}
\label{Sec:model}

\begin{figure}[tb]
\centerline{
\includegraphics[width=75mm]{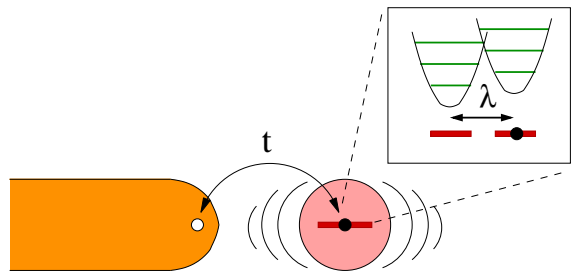}
}\vspace{0pt}
\caption{(Color online) The physical system under
         consideration. A localized level with energy
         $\epsilon^{}_d$ is tunnel coupled with
         amplitude $t$ to a band of spinless electrons,
         and is simultaneously coupled by a
         dimensionless displacement coupling $\lambda$
         to a single vibrational mode of frequency
         $\omega_0$. Depending on the valence of the
         level, the vibrational mode experiences a
         shifted harmonic potential with a relative
         displacement of $\Delta b = \lambda$.}
\label{Fig:System}
\end{figure}

The system under consideration is shown schematically
in Fig.~\ref{Fig:System}. It consists of a single
localized electronic level $d^{\dagger}$ with energy
$\epsilon^{}_d$, tunnel coupled to a continuous band
of noninteracting spinless electrons which we denote by
$c^{\dagger}_{k}$. The level is simultaneously coupled
by displacement to a localized vibrational mode
(phonon), as modeled by the Hamiltonian
\begin{eqnarray}
{\cal H} &=&
         \sum_{k} \epsilon_k c^{\dagger}_{k} c^{}_{k}
         + \frac{t}{\sqrt{N}}
               \sum_{k} \left \{
                                 d^{\dagger} c^{}_{k}
                                 + {\rm H.c.}
                        \right \}
         + \epsilon^{}_d d^{\dagger} d
\nonumber \\
      && + \omega_{0} b^{\dagger} b
         + \lambda \omega_0
           \left (
                   d^{\dagger} d - N_0
           \right )
           \left (
                   b^{\dagger} + b
           \right ) .
\label{H-ep}
\end{eqnarray}
Here, $b^{\dagger}$ creates a local Einstein phonon
that oscillates with frequency $\omega_0$, $t$ is the
tunneling matrix element between the level and the
Wannier state closest to the molecule, and $N$ is the
number of lattice sites. The dimensionless coupling
$\lambda$ measures the relative displacement of the
vibrational mode between the configurations where
the level is empty and occupied. It serves as
a faithful measure for the strength of the
electron-phonon coupling, with $\lambda \ll 1$
($1 \ll \lambda$) corresponding to weak (strong)
interactions. The parameter $N_0$ can be thought of
as fixing the reference charge of the level. It can
be formally eliminated by shifting the bosonic mode
according to $\hat{B} = b - \lambda N_0$, which
has the effect of renormalizing the level energy
from its bare value $\epsilon^{}_d$ to
$\tilde{\epsilon}^{}_d$ with
\begin{equation}
\tilde{\epsilon}^{}_d = \epsilon^{}_d
                  + 2 N_0 \lambda^2 \omega_0 .
\label{e_d-renormalized}
\end{equation}
The conversion from $b$ to $B$ also generates
the constant term $-N_0 \lambda^2 \omega_0$,
which uniformly shifts the spectrum of the
Hamiltonian. Although the inclusion of $N_0$ adds
no richness to the thermodynamics of the model, it
provides a useful tuning parameter for exploring
the low-energy state of the system, as will be
demonstrated later on.

In the following we shall consider a particle-hole
symmetric band, namely, it is assumed that the wave
numbers can be grouped into distinct pairs $k$ and
$k'$, such that $\epsilon_{k'} = -\epsilon_k$ for
each pair of momenta. It is easy to verify that the
combined transformation
\begin{equation}
d^{\dagger} \to d , \;\;\;
c^{\dagger}_{k} \to -c^{}_{k'} , \;\;\;
{\rm and} \;\;\; b \to -b - \lambda (1 - 2N_0)
\label{transformation}
\end{equation}
leaves the Hamiltonian of Eq.~(\ref{H-ep})
unchanged under these terms, apart from the
substitution~\cite{comment-on-shift-I}
\begin{equation}
\epsilon^{}_d \to
      \tilde{\epsilon}^{}_d =
      2 \lambda^2 \omega_0 (1 - 2N_0) - \epsilon^{}_d .
\end{equation}
It therefore suffices to study the domain
$\epsilon^{}_{d} \geq \epsilon_{d}^{\ast}$ with
\begin{equation}
\epsilon_{d}^{\ast} =
         \lambda^2 \omega_0 (1 - 2N_0) ,
\end{equation}
while the complementary domain $\epsilon^{}_{d} <
\epsilon_{d}^{\ast}$ is accessible via the particle-hole
transformation of Eq.~(\ref{transformation}).
In particular, occupancy of the level
$n_d (\epsilon^{}_d) = \langle d^{\dagger} d \rangle$
obeys the symmetry relation
\begin{equation}
n_d(\epsilon_d^{\ast} - \Delta \epsilon_d^{}) = 1 -
    n_d(\epsilon_d^{\ast} + \Delta \epsilon_d^{}) ,
\label{n_d-symmetry}
\end{equation}
independent of all other model parameters, the
temperature included.

Other than the conduction-electron bandwidth $D$,
the Hamiltonian of Eq.~(\ref{H-ep}) features four
basic energy scales. These include the bare
vibrational frequency $\omega_0$, the polaronic
shift $E_{\rm p} = \lambda^2 \omega_0$, the detuning
energy $\Delta \epsilon_{d}^{} = \epsilon_{d}^{} -
\epsilon_{d}^{\ast}$, and the hybridization width
$\Gamma = \pi \rho_0 t^2$. Here $\rho_0$ is the
conduction-electron density of states at the Fermi
level.~\cite{comment-on-DOS} Of particular interest
is the point
$\Delta \epsilon^{}_d = 0$, when the (renormalized)
level lies at resonance with the Fermi energy.
The relevant low-energy scale in the problem
is conveniently defined~\cite{BSZ08} in this case
from the zero-temperature charge susceptibility
evaluated at $\epsilon_{d}^{\ast}$:
\begin{equation}
\Gamma_{\rm eff} = 
       \frac{1}{\pi \chi_{\rm c}} ,
\label{G_eff-def}
\end{equation}
with
\begin{equation}
\chi_c = \left.
               -\frac{d n_{d}}{d\epsilon^{}_{d}}
         \right |_{\epsilon_d^{} = \epsilon_d^{\ast}} .
\label{chi_c-def}
\end{equation}

For $\lambda = 0$, the low-energy scale $\Gamma_{\rm eff}$
so defined coincides with the bare hybridization width
$\Gamma$. Various thermodynamic properties associated
with the level, e.g., its occupancy and its contribution
to the electronic specific heat, reduce in the wide-band
limit to exclusive functions of $\epsilon^{}_d/\Gamma$
and $T/\Gamma$, where $T$ is the temperature. A nonzero
$\lambda$ modifies this picture both qualitatively and
quantitatively. For example, occupancy of the level at
$T = 0$ is no longer given for large $\lambda$ by an
exclusive function of
$\Delta \epsilon^{}_d/\Gamma_{\rm eff}$, but rather
depends, as we shall show, on vastly different energy
scales. Our goal is to conduct a systematic study of
all coupling regimes for nonzero $\lambda$, focusing
primarily on large $\lambda$. To this end, we shall
combine analytical considerations with Wilson's
renormalization-group (NRG) method.~\cite{Wilson75,BCP08}

\section{Limiting cases}
\label{Sec:limits}

We begin with a preliminary discussion of certain
limits where analytic insight can be gained. These
include the traditional adiabatic and antiadiabatic
limits, the proposed extended antiadiabatic regime
where $\omega_0 < \Gamma$ but 
$\Gamma^{}_{\rm eff} \ll \omega_0$, and the
limit of large detuning, $\Gamma \ll |\epsilon_d|$.

\subsection{Adiabatic limit}


\begin{figure}[tb]
\centerline{
\includegraphics[width=75mm]{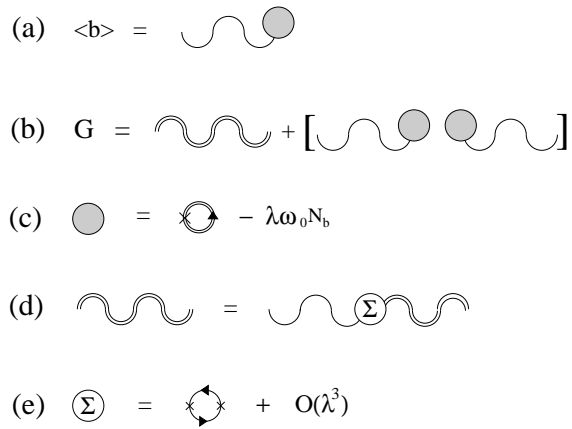}
}\vspace{0pt}
\caption{Diagrammatic representation of (a) the phononic
         expectation value $\langle b \rangle$ and (b)
         the phononic Green's function $G(i \nu_n)$.
         Here, single (double) wiggly lines represent
         the bare (dressed) connected phonon propagator
         $\tilde{G}(i \nu_n)$, whereas single (double)
         lines with an arrow denote the bare (dressed)
         $d$-electron propagator (bare/dressed with
         respect to $\lambda$; both electronic propagators
         are dressed with respect to $t$ alone). The
         connected phonon
         propagator $\tilde{G}(i \nu_n)$ obeys a standard
         Dyson equation (d), with a matrix self-energy
         $\hat{\Sigma}$ whose second-order expansion in
         $\lambda$ is specified in panel (e). Interaction
         vertices are marked by a cross.}
\label{fig:Diagrams}
\end{figure}

Commencing with the limit of small $\lambda$, we apply
ordinary perturbation theory in the electron-phonon
coupling $\lambda$, postponing for the moment the
question of its range of validity. The basic quantity of
interest is the Matsubara phonon propagator
\begin{equation}
G(i \nu_n) =
    \left [
            \begin{array}{cc}
                G_{b b^{\dagger}}(i \nu_n) &
                G_{b b}(i \nu_n) \\ \\
                G_{b^{\dagger} b^{\dagger}}(i \nu_n) &
                G_{b^{\dagger} b}(i \nu_n)
            \end{array}
    \right ] ,
\end{equation}
where $\nu_n = 2 \pi n/\beta$ are the bosonic Matsubara
frequencies, $\beta = 1/k_{\rm B} T$ is the reciprocal
temperature, and
\begin{equation}
G_{A B}(i \nu_n) =
         -\int_{0}^{\beta}
               \langle T_{\tau} \hat{A}(\tau)
                                \hat{B}(0) \rangle
               e^{i \nu_n \tau} d\tau .
\end{equation}
Using the diagrams specified in Fig.~\ref{fig:Diagrams},
one obtains the formally exact relations
\begin{equation}
\langle b \rangle = -\lambda
                     \left [
                             \langle d^{\dagger} d \rangle
                             - N_0
                     \right ]
\end{equation}
and
\begin{equation}
G_{\alpha \gamma}(i \nu_n) = 
          \tilde{G}_{\alpha \gamma}(i \nu_n)
          - \delta_{n, 0} \beta \lambda^2 
                    \left [
                            \langle d^{\dagger} d \rangle
                            - N_0
                    \right ]^2 ,
\end{equation}
where $\tilde{G}(i \nu_n)$ is the connected phonon
propagator, defined as the sum of all connected phonon
diagrams. The connected propagator has the conventional
representation $\tilde{G}^{-1} = [\tilde{G}^{(0)}]^{-1}
- \hat{\Sigma}$, where
\begin{equation}
\left [
        \tilde{G}^{(0)}(i \nu_n)
\right ]^{-1} =
     \left (
             \begin{array}{cc}
                   i\nu_n - \omega_0 & 0 \\ \\
                   0 & -i\nu_n - \omega_0
             \end{array}
     \right )
\end{equation}
is the unperturbed phonon Green's function and
$\hat{\Sigma}$ is the self-energy matrix. Since
the electron-phonon interaction in Eq.~(\ref{H-ep})
involves only the combination $b + b^{\dagger}$,
the self-energy matrix takes the general form
\begin{equation}
\hat{\Sigma}(i \nu_n) =
     \sigma(i \nu_n)
     \left (
             \begin{array}{cc}
                   1 & 1 \\
                   1 & 1
             \end{array}
     \right ) ,
\end{equation}
which depends on a single scalar function
$\sigma(i \nu_n)$. Settling with second order in
$\lambda$ and analytically continuing to real
frequencies, $\sigma(\omega + i\eta)$ is given for
$T \to 0$ in the wide-band limit by
\begin{equation}
\sigma(\omega + i\eta) =
      \frac{\lambda^2 \omega_0^2}{2 i \pi}
      \left [
              I_{-}(\omega) - I_{-}^{\ast}(-\omega)
              - I_{+}(\omega) + I_{+}^{\ast}(-\omega)
      \right ] ,
\end{equation}
where
\begin{equation}
I_{+}(\omega) = \frac{1}{\omega}
                \left [
                        \ln (-\epsilon_d + i\Gamma)
                        - \ln (\omega - \epsilon_d
                               + i\Gamma)
                \right ]
\end{equation}
and
\begin{equation}
I_{-}(\omega) = \frac{1}{\omega + 2i \Gamma}
                \left [
                        i\pi +
                        \ln (\epsilon_d + i\Gamma)
                        - \ln (\omega - \epsilon_d
                               + i\Gamma)
                \right ] .
\end{equation}
Focusing on the resonance condition
$\epsilon^{}_d = 0$ and expanding in powers of
the frequency,~\cite{comment-on-resonance}
$\sigma(\omega + i\eta)$ reads as
\begin{equation}
\sigma(\omega + i\eta) = 
              -\frac{\lambda^2 \omega_0^2}{\pi \Gamma}
              -i \omega
               \frac{\lambda^2 \omega_0^2}{\pi \Gamma^2}
              + {\cal O}(\omega^2) ,
\end{equation}
resulting in
\begin{equation}
\tilde{G}^{-1}(\omega + i \eta) =
  \left [
          \begin{array}{cc}
                (1\!+\!i\gamma) \omega\!-\!\omega_0\!+\!A &
                i\gamma \omega\!+\!A \\ \\
                i\gamma \omega\!+\!A &
                (i\gamma\!-\!1) \omega\!-\!\omega_0\!+\!A
          \end{array}
  \right ]
\end{equation}
with $\gamma = \lambda^2 \omega_0^2/(\pi \Gamma^2)$
and $A = \lambda^2 \omega_0^2/(\pi \Gamma)$. The poles
of $\tilde{G}(z)$ can now be identified with the zeros
of ${\rm det} \{ \tilde{G}^{-1}(z) \}$, which yields
\begin{equation}
z_{\pm} = -i \frac{\omega_0}{\pi}
           \left (
                   \frac{\lambda \omega_0}{\Gamma}
           \right )^2
        \pm \omega_0
            \sqrt{1 - \frac{1}{\pi^2}
                      \left (
                              \frac{\lambda \omega_0}
                                   {\Gamma}
                      \right )^4
                    - \frac{2}{\pi} \frac{E_{\rm p}}
                                         {\Gamma}
                 } \,.
\label{z_pm}
\end{equation}

Several conclusions can be drawn from Eq.~(\ref{z_pm}).
First, there are two distinct parameters that
control the perturbative expansion: $E_p/\Gamma$
and $\lambda \omega_0/\Gamma$. Depending on the
magnitude of $\lambda$ either parameter can be the
largest, with $E_p/\Gamma$ dominating for $1 \ll \lambda$.
Second, both parameters must be small in order for
a weakly damped phonon mode to persist. A crossover
to an overdamped phonon occurs as soon as either of
these parameters becomes of order unity, signaling
a qualitative change in the underlying physics
and the breakdown of perturbation theory.
Third, coupling to the electronic level softens
the phonon frequency according to
\begin{equation}
\frac{\omega_{\rm eff}}{\omega_0}
          = \sqrt{1 - \frac{1}{\pi^2}
                      \left (
                              \frac{\lambda \omega_0}
                                   {\Gamma}
                      \right )^4
                    - \frac{2}{\pi} \frac{E_{\rm p}}
                                         {\Gamma}
                 } \,.
\label{omega_eff}
\end{equation}
For $1 \ll \lambda$ we therefore conclude that the
perturbative physics of the adiabatic limit breaks
down as soon as the polaronic shift exceeds $\Gamma$.


\subsection{Antiadiabatic limit}
\label{Sec:Anti-A}

When $\Gamma$ is sufficiently small as compared to
$\omega_0$, all phonon excitations are frozen out
as the temperature is decreased below $\omega_0$.
Only fermionic excitations remain active at such
low energies, reflecting the fact that the relevant
electronic motion is far slower than the phonon
vibrations. To derive the effective low-energy
Hamiltonian for $\Delta \epsilon^{}_d = 0$, it
is useful to first apply the Lang-Firsov
transformation~\cite{LF62}
${\cal H}' = \hat{U}^{\dagger} {\cal H} \hat{U}$
with $\hat{U} = e^{-\lambda (b^{\dagger} + b)
(d^{\dagger} d - N_0)}$, which converts
Eq.~(\ref{H-ep}) into
\begin{equation}
{\cal H}' =
         \sum_{k} \epsilon_k
                  c^{\dagger}_{k} c^{}_{k}
         + \frac{t}{\sqrt{N}}
               \sum_{k} \left \{
                                 e^{\lambda
                                    (b^{\dagger} - b)}
                                 d^{\dagger} c^{}_{k}
                                 + {\rm H.c.}
                        \right \}
         +\omega_0 b^{\dagger} b .
\label{H-LF}
\end{equation}
Here we have set $\Delta \epsilon^{}_d = 0$ and omitted
the global energy shift $\Delta E = -\lambda^{2}
\omega_0 N_{0}^{2}$. The effect of the Lang-Firsov
transformation is to eliminate the displacement
interaction term at the expense of attaching the
exponentials $e^{\pm \lambda (b^{\dagger} - b)}$ to the
tunneling amplitude $t$. The transformed Hamiltonian
is invariant under the particle-hole transformation
\begin{equation}
d^{\dagger} \to d , \;\;\;
c^{\dagger}_{k} \to -c^{}_{k'} , \;\;\;
{\rm and} \;\;\; b \to -b ,
\label{transformation-II}
\end{equation}
which comes in place of Eq.~(\ref{transformation}).

Provided $\Gamma$ is small enough, the (transformed)
phonon is frozen in its unperturbed ground state
configuration $b^{\dagger} b = 0$ at energies well
below $\omega_0$. The effective Hamiltonian at such
low energies is purely fermionic, and may generally
contain all possible local Hamiltonian terms that are
invariant under the particle-hole transformation
$d^{\dagger} \to d$, $c^{\dagger}_{k} \to -c^{}_{k'}$.
A rigorous derivation of the effective low-energy
Hamiltonian requires a systematic integration of
all high-energy excitations, including a successive
elimination of all discrete phonon excitations. This
turns out to be a difficult task. Nevertheless, for
$\Gamma \ll \omega_0$ one can settle with a
single-step elimination of all excitation energies
exceeding $\omega_0$ using a Schrieffer-Wolff-type
transformation.~\cite{SW66} Deferring the details
of the derivation to Appendix~\ref{app:SWT}, we
quote here only the end result.

To second order in $t$, one is left with an effective
interacting resonant-level model~\cite{VF78,Schlottmann82}
(IRLM) of the form
\begin{align}
{\cal H}_{\rm eff}
        & =
             \sum_{k}^{\rm restric} \epsilon_k
                     c^{\dagger}_{k} c^{}_{k}
             + \frac{t_{\rm eff}}{\sqrt{N}}
                    \sum_{k}^{\rm restric}
                        \left \{
                                 d^{\dagger} c^{}_{k}
                                 + {\rm H.c.}
                        \right \}
\label{H-eff} \\
         & + \frac{U_{\rm eff}}{N}
             \bigl (
                     d^{\dagger} d - 1/2
             \bigr )
             \sum_{k, k'}^{\rm restric}
                   :\! c^{\dagger}_{k} c^{}_{k'}\!\!: .
\nonumber
\end{align}
Here, $:\! c^{\dagger}_{k} c^{}_{k'}\!\!: =
c^{\dagger}_{k} c^{}_{k'} - \theta(-\epsilon_k)
\delta_{k, k'}$ stands for normal ordering with
respect to the filled Fermi sea, while the symbol
$\sum_{k}^{\rm restric}$ comes to indicate that
the summation over $k$ is restricted to momenta
such that $|\epsilon_k| < \omega_0$ ($N' < N$
being the number of such $k$ points). The coupling
constants entering Eq.~(\ref{H-eff}) are given by
\begin{equation}
t_{\rm eff} = t e^{-\lambda^2/2}
\label{t_eff}
\end{equation}
and
\begin{equation}
U_{\rm eff} = \frac{2 t^2}{\omega_0}
              e^{-\lambda^2}
              \bigl [
                      {\rm Ei}(\lambda^2) -
                      2\ln (\lambda) - \gamma
              \bigr ] ,
\label{U_eff}
\end{equation}
where ${\rm Ei}(x)$ is the exponential integral
function~\cite{AS-Ei} and $\gamma \approx 0.5772$ is
Euler's constant. In the limits where $\lambda$ is
either small or large as compared to one,
$U_{\rm eff}$ takes the simplified forms
\begin{equation}
U_{\rm eff} \approx
            \left \{
            \begin{array}{cc}
                  2 t^2 \lambda^2/\omega_0 , \;\; &
                  \lambda \ll 1 , \\ \\
                  2 t^2/(\lambda^2 \omega_0) , \;\; &
                  1 \ll \lambda ,
            \end{array}
            \right .
\label{U_eff-weak-t}
\end{equation}
as follows from the corresponding asymptotes of the
exponential integral function.

One can borrow at this
point known results for the IRLM in order to extract
the renormalized level width $\Gamma_{\rm eff}$.
Specifically, it is known from perturbative
renormalization-group calculations~\cite{BVZ07} that
\begin{equation}
\Gamma_{\rm eff} = D_{\rm eff}
                 \left (
                         \frac{\tilde{\Gamma}}
                              {D_{\rm eff}}
                 \right )^{1/(1 + 2 \rho_0 U_{\rm eff})} ,
\end{equation}
where $D_{\rm eff}$ is the effective bandwidth
and $\tilde{\Gamma} = \pi \rho_0 t_{\rm eff}^2$.
Inserting Eq.~(\ref{t_eff}) for $t_{\rm eff}$ and
setting $D_{\rm eff} \sim \omega_0$ yields
\begin{equation}
\Gamma_{\rm eff} = \omega_0
                 \left (
                         \frac{\Gamma}{\omega_0}
                         e^{-\lambda^2}
                 \right )^{1/(1 + 2 \rho_0 U_{\rm eff})} .
\label{G_eff}
\end{equation}
It should be stressed that Eqs.~(\ref{t_eff}),
(\ref{U_eff}), and (\ref{G_eff}) are restricted to
$\Gamma \ll \omega_0$, when the exponent that appears
in Eq.~(\ref{G_eff}) hardly deviates from one. As we
show below, the reduction to an effective IRLM is far
more general, though, and applies to all parameter
regimes where $\Gamma_{\rm eff} \ll \omega_0$.

\subsection{Extended antiadiabatic regime}
\label{Sec:Extended-AAL}

Once $\Gamma$ exceeds $\omega_0$, one looses
the hierarchy of energy scales underlying the
derivation of the effective low-energy Hamiltonian
of Eq.~(\ref{H-eff}). Nevertheless, the general
form of the low-energy Hamiltonian can still
be deduced for $\Gamma_{\rm eff} \ll \omega_0$
on physical grounds. Whenever
$\Gamma_{\rm eff} \ll \omega_0$, the renormalized
electronic motion remains sufficiently slow for
the phonon to efficiently respond to the individual
tunneling events. Hence, one can expect a rather
well-defined bosonic mode to persist, with a
modified frequency that remains close in magnitude
to $\omega_0$. As in the strict antiadiabatic
limit, the bosonic mode is frozen in its
ground-state configuration as the
temperature is lowered below $\omega_0$, leaving
only purely fermionic excitations active at such
low energies. With increasing $\Gamma$, the
microscopic details of the bosonic mode and its
associated polaron will progressively deviate
from their $\Gamma \to 0$ forms, yet the
effective low-energy Hamiltonian remains purely
fermionic as long as $\Gamma_{\rm eff} \ll \omega_0$.
As we argue below, the general form of the resulting
Hamiltonian is essentially dictated by symmetry
considerations.

To illustrate this point, let us begin with the
transformed Hamiltonian of Eq.~(\ref{H-LF}).
As indicated above, the effective fermionic
Hamiltonian may generally contain all possible
local Hamiltonian terms that are invariant
under the particle-hole transformation
$d^{\dagger} \to d$, $c^{\dagger}_{k} \to -c^{}_{k'}$.
Of all possible terms in this category, only the
local tunneling term $t$ is relevant in the
renormalization-group sense, while the local
contact interaction
\begin{equation}
U \bigl (
          d^{\dagger} d - 1/2
  \bigr )
  \sum_{k, k'} :\! c^{\dagger}_{k} c^{}_{k'}\!\!:
\end{equation}
is marginal. All other local terms permitted by
symmetry are formally irrelevant, leaving us
with the effective IRLM of Eq.~(\ref{H-eff}).
A nonzero $\Delta \epsilon^{}_d$ relaxes the
requirement of particle-hole symmetry, which
augments the Hamiltonian of Eq.~(\ref{H-eff})
with two additional terms:
\begin{equation}
\epsilon_{\rm eff} \, d^{\dagger} d 
      + \frac{V_{\rm eff}}{N}
             \sum_{k, k'}^{\rm restric}
                   :\! c^{\dagger}_{k} c^{}_{k'}\!\!: .
\end{equation}

Note that although the form of Eq.~(\ref{H-eff}) is
dictated by symmetry considerations, these
considerations alone do not suffice to fix the values
that the couplings $t_{\rm eff}$ and $U_{\rm eff}$
acquire. These couplings must generally be extracted
from the low-energy spectrum of the original
electron-phonon Hamiltonian, as will be done later
on using the NRG. As we shall show, the condition
$\Gamma_{\rm eff} \ll \omega_0$ encompasses for
$1 \ll \lambda$ all values of $\Gamma$ up to
$\Gamma \sim 0.4 E_p$. Furthermore, $U_{\rm eff}$
remains well described by Eq.~(\ref{U_eff}) for most
of this range, whereas $t_{\rm eff}$ rapidly exceeds
Eq.~(\ref{t_eff}) as soon as $\Gamma$ approaches
$\omega_0$.

\subsection{Large detuning}

One limit where perturbation theory in $t$ is
guaranteed to apply is that of large detuning,
$\Delta \epsilon^{}_d \gg \Gamma$ (equivalently
$-\Delta \epsilon^{}_d \gg \Gamma$). In this case
$\Delta \epsilon^{}_d$ is sufficiently large
to assure that the electronic level remains
nearly empty at zero temperature, providing
a suitable starting point for a perturbative
expansion in $t$. In the following we focus on
the zero-temperature occupancy of the level,
$n_d(\epsilon^{}_d)$, which is conveniently
computed from the derivative of the ground-state
energy with respect to $\epsilon^{}_d$.

To obtain the correction to the ground-state
energy it is useful to start from the transformed
Hamiltonian of Eq.~(\ref{H-LF}), which is augmented
for $\Delta \epsilon^{}_d \neq 0$ by the Hamiltonian
term $\Delta \epsilon^{}_d d^{\dagger} d$. For $t = 0$
and $\Delta \epsilon^{}_d > 0$, the ground state
of ${\cal H}'$ is given by the product state of the
filled Fermi sea with an empty level and the empty
phonon state. To second order in $t$ the ground-state
energy acquires the correction
\begin{equation}
\delta E_{\rm gs} =
    \frac{\Gamma}{\pi} e^{-\lambda^2}
    \sum_{n = 0}^{\infty}
         \frac{\lambda^{2n}}{n!}
         \ln
         \left(
                 \frac{\Delta\epsilon^{}_d + n\omega_0}
                      {\Delta\epsilon^{}_d + D + n\omega_0}
         \right ) ,
\label{delta-E_gs}
\end{equation}
where we have assumed a symmetric rectangular
density of states for the conduction electrons:
$\rho(\epsilon) = \rho_0 \theta (D - |\epsilon|)$.
Straightforward differentiation of Eq.~(\ref{delta-E_gs})
with respect to $\epsilon_d$ yields then the level
occupancy
\begin{align}
n_d(\Delta \epsilon^{}_d > 0) =
      \frac{\Gamma}{\pi} e^{-\lambda^2} 
      \sum_{n = 0}^{\infty}
            \frac{\lambda^{2n}}{n!} &
            \left (
                    \frac{1}
                         {\Delta \epsilon^{}_d + n\omega_0}
            \right.
\label{n_d-perturbative} \\
          & \;\;
            \left.
                    - \frac{1}
                           {D + \Delta \epsilon^{}_d + n\omega_0}
            \right ) ,
\nonumber
\end{align}
which properly reduces for $\lambda \to 0$ to the
noninteracting result
\begin{equation}
n_d(\epsilon^{}_d > 0) = \frac{\Gamma}{\pi}
      \left (
              \frac{1}{\Delta \epsilon^{}_d}
              - \frac{1}{D + \Delta \epsilon^{}_d}
      \right ) .
\end{equation}
In the wide-band limit, $D \to \infty$, the
second term drops out in the parentheses of
Eq.~(\ref{n_d-perturbative}).

Note that Eq.~(\ref{n_d-perturbative}) was
derived under the strict condition that
$\Delta \epsilon^{}_d \gg \Gamma$. Below we
present NRG results that suggest a broader range
of validity of Eq.~(\ref{n_d-perturbative}), down
to $\Delta \epsilon^{}_d \sim \Gamma_{\rm eff}$.
Since $\Gamma_{\rm eff} \ll \Gamma$ for
$1 \ll \lambda$ and $\Gamma \ll E_p$, this
implies a far greater range of validity of
perturbation theory in $t$.

\subsection{Phonon distribution function}
\label{Sec:P(n)-analytic}

Our discussion thus far was restricted to
electronic properties. Another quantity of
interest is the phonon distribution function
\begin{equation}
P(n) = \bigl <
               |n \rangle \langle n |
       \bigr > ,
\label{P(n)-def}
\end{equation}
which contains direct information on the state
of the phonon. The phonon distribution function
has distinct characteristic forms in the extreme
adiabatic and antiadiabatic limits, which we next
derive. The transition between these two limiting
forms indicates the undressing of the polaron upon
going from the antiadiabatic to the adiabatic
regime.

In the perturbative adiabatic regime, the phonon
is too slow to respond to the successive tunneling
events, hence it samples only the time-averaged
occupancy of the level. From the standpoint of
the phonon it therefore experiences the effective
Hamiltonian
\begin{equation}
{\cal H}_{\rm phonon} = 
      \omega_{0} b^{\dagger} b
      + \lambda \omega_0 ( n_d - N_0)
           \left (
                   b^{\dagger} + b
           \right ) ,
\end{equation}
describing the average displacement $\lambda \to
\bar{\lambda} = \lambda ( n_d - N_0)$. At
$T = 0$ the phonon is thus frozen in the coherent
state $b |\bar{\lambda} \rangle = -\bar{\lambda}
|\bar{\lambda} \rangle$, resulting in
\begin{equation}
P_{\rm adiabatic}(n) =
      \frac{\bar{\lambda}^{2n}}{n!}
      e^{-\bar{\lambda}^2} .
\label{P(n)-AL}
\end{equation}
For the particular case where $\Delta \epsilon^{}_d = 0$
and $N_0 = 0$, one has that $\bar{\lambda} = \lambda/2$.

In the antiadiabatic limit, it is advantageous to
consider first the ground state $|\psi'_{\rm gs} \rangle
= \hat{U}^{\dagger}|\psi_{\rm gs}\rangle$ of
the transformed Hamiltonian ${\cal H}'$ of
Eq.~(\ref{H-LF}). Here $\hat{U} = e^{-\lambda
(b^{\dagger} + b) (d^{\dagger} d - N_0)}$ is the
canonical transformation relating ${\cal H}$
and ${\cal H}'$. As discussed in
Sec.~\ref{Sec:Anti-A}, for $\Gamma \ll \omega_0$
the transformed phonon is effectively frozen
at $T = 0$ in its unperturbed ground state
$b^{\dagger} b = 0$. This means that
$|\psi'_{\rm gs} \rangle$ is well approximated
by a product state of the form
\begin{equation}
|\psi'_{\rm gs} \rangle = |\phi \rangle_{\rm el}
                \otimes |n = 0 \rangle_{\rm ph} ,
\end{equation}
where $|\phi \rangle_{\rm el}$ pertains to the
electronic degrees of freedom. The ground state
of the original Hamiltonian $|\psi_{\rm gs}\rangle$
can now be obtained by applying the transformation
$\hat{U}$ to $|\psi'_{\rm gs}\rangle$, resulting
in
\begin{equation}
|\psi'_{\rm gs} \rangle =
       |\phi_0 \rangle_{\rm el} \otimes
       |\lambda_0 \rangle_{\rm ph} 
       + |\phi_1 \rangle_{\rm el} \otimes
         |\lambda_1 \rangle_{\rm ph} .
\end{equation}
Here $|\phi_0 \rangle_{\rm el}$ and
$|\phi_1 \rangle_{\rm el}$, respectively, are
the projections of the electronic state
$|\phi \rangle_{\rm el}$ onto the $\hat{n}_d = 0$
and $\hat{n}_d = 1$ subspaces, while
$|\lambda_0 \rangle_{\rm ph}$ and
$|\lambda_1 \rangle_{\rm ph}$ are the phononic
coherent states with $\lambda_0 = \lambda N_0$
and $\lambda_1 = \lambda (N_0 - 1)$. Accordingly,
the phonon distribution function assumes the form
\begin{equation}
P_{\rm anti-adiabatic}(n) =
      (1 - n_d)
      \frac{\lambda_0^{2n}}{n!}
           e^{-\lambda_0^2}
      + n_d \frac{\lambda_1^{2n}}{n!}
                 e^{-\lambda_1^2} .
\end{equation}
In particular, for $N_0 = \Delta \epsilon^{}_d = 0$,
\begin{equation}
P_{\rm anti-adiabatic}(n) =
          \frac{1}{2}
          \left [
                  \delta_{n, 0} +
                  \frac{\lambda^{2n}}{n!} e^{-\lambda^2}
          \right ] .
\label{P(n)-AAL}
\end{equation}

\section{Systematic study of all coupling regimes}
\label{Sec:NRG-study}

\subsection{The Numerical renormalization group}
\label{Sec:NRG}

To treat the Hamiltonian of Eq.~(\ref{H-ep}) for arbitrary
coupling strengths, we resort to Wilson's numerical
renormalization-group method~\cite{Wilson75,BCP08}
(NRG). The NRG is a powerful tool for accurately
calculating equilibrium properties of arbitrarily complex
quantum impurities. Originally devised for treating
the single-channel Kondo Hamiltonian,~\cite{Wilson75}
this nonperturbative approach was successfully
applied over the years to numerous impurity models and
setups.~\cite{BCP08} At the heart of the approach is
a logarithmic energy discretization of the conduction
band about the Fermi energy, controlled by the
discretization parameter $\Lambda > 1$. Using an
appropriate unitary transformation,~\cite{Wilson75}
the conduction band is mapped onto a semi-infinite
chain with the impurity coupled to its open end. The
$N$th link along the chain represents an exponentially
decreasing energy scale $D_N \sim \Lambda^{-N/2}$, with
the continuum limit recovered for $\Lambda \to 1^{+}$.
The full Hamiltonian of Eq.~(\ref{H-ep}) is thus
recast as a double limit of a sequence of
dimensionless NRG Hamiltonians:
\begin{equation}
{\cal H} = \lim_{\Lambda \rightarrow 1^+}
           \lim_{N \rightarrow \infty}
           \left\{
                 D_{\Lambda} \Lambda^{-(N-1)/2} {\cal H}_{N}
           \right\} ,
\end{equation}
with $D_{\Lambda} = D(1 + \Lambda^{-1})/2$ and
\begin{eqnarray}
{\cal H}_N &=& \Lambda^{\frac{N-1}{2}}
               \left [
                       \tilde{\epsilon}_{d} d^{\dagger} d
                       + \tilde{\omega}_{0} b^{\dagger} b
                       + \tilde{t}
                         \left \{
                                  f^{\dagger}_{0} d
                                + d^{\dagger} f_0
                         \right \}
               \right.
\nonumber \\
&&+
                         \lambda
                         \tilde{\omega}_{0}
                         \left (
                                 d^{\dagger} d - N_0
                         \right )
                         \left (
                                 b^{\dagger} + b
                         \right )
\nonumber \\
&&+
               \left.
                     \sum_{n = 0}^{N-1}
                           \Lambda^{-\frac{n}{2}} \xi_{n}
                           \left \{
                                    f^{\dagger}_{n+1}
                                    f_{n}
                                    + {\rm H.c.}
                           \right \}
               \right ] .
\label{H-NRG}
\end{eqnarray}
Here, 
$\tilde{\epsilon}_{d} = \epsilon_{d}/D_{\Lambda}$ and
$\tilde{\omega}_{0} = \omega_{0}/D_{\Lambda}$ are the
dimensionless energy level and vibrational frequency,
respectively, while $\tilde{t}$ is related to the
tunneling matrix element through
\begin{equation}
\tilde{t} = \sqrt{A_{\Lambda}} \frac{t}{D_{\Lambda}} .
\label{t-tilde}
\end{equation}
The coefficient
\begin{equation}
A_{\Lambda} = \frac{\Lambda + 1}{2 (\Lambda - 1)}
              \ln \Lambda
\end{equation}
is required to account for the energy discretization
used in the NRG,~\cite{KWW80a} and can be viewed as
accelerating the convergence to the $\Lambda \to 1^{+}$
limit. The prefactor $\Lambda^{(N-1)/2}$ that appears
in Eq.~(\ref{H-NRG}) comes to ensure that the low-lying
excitations of ${\cal H}_N$ are of order one for all
$N$.

Physically, the shell operator $f^{\dagger}_{0}$
represents the local conduction-electron state to
which the level is directly coupled by tunneling.
The subsequent shell operators $f^{\dagger}_{n}$
correspond to wave packets whose spatial extent
about the level grows roughly as $\Lambda^{n/2}$.
Details of the band are encoded in the hopping
coefficients $\xi_{n}$, obtained from suitable integrals
of the density of states.~\cite{BPH97} Throughout the
paper we assume a relativistic dispersion relation,
corresponding to the symmetric rectangular density of
states $\rho(\epsilon) = \rho_0 \theta ( D - |\epsilon|)$
with $\rho_0 = 1/(2 D)$.
This simplified form of $\rho(\epsilon)$ affords an
explicit analytical expression for
$\xi_{n}$,~\cite{Wilson75} which rapidly approaches
one with increasing $n$.

A key ingredient of the NRG is the separation of energy
scales along the Wilson chain, which enables an iterative
diagonalization of the sequence of finite-size
Hamiltonians ${\cal H}_N$. Starting from a core
cluster that consists of the local degrees of freedom
$d^{\dagger}$, $b^{\dagger}$, and $f^{\dagger}_0$, the
Wilson chain is successively enlarged by adding one
site at a time. Diagonalization of ${\cal H}_{N + 1}$
proceeds from the knowledge of the spectrum of
${\cal H}_N$ by means of the NRG transformation
\begin{equation}
{\cal H}_{N+1} = \sqrt{\lambda}{\cal H}_N
               + \xi_{N}
                 \left \{
                         f^{\dagger}_{N+1} f_{N}
                         + {\rm H.c.}
                 \right \} .
\label{NRG_iteration}
\end{equation}
In this manner, one can track the evolution of the
finite-size spectrum as a function of $N$. The approach
to a fixed point is signaled by a limit cycle of the
NRG transformation, with ${\cal H}_{N + 2}$ and
${\cal H}_{N}$ sharing the same low-energy spectrum.

The above procedure could, in principle, be applied
to any set of hopping matrix elements along the chain.
However, practical considerations prove far more
restrictive, as it is numerically impossible to keep
track of the exponential growth of the Hilbert space
with increasing $N$. In practice only a limited number
of states can be retained at the
conclusion of each NRG iteration, which is where
the separation of scales along the chain comes
into play. Due to the exponential decrease of
the hopping terms, one can settle with retaining
only the lowest $N_{\rm s}$ eigenstates of
${\cal H}_N$ when constructing the low-energy
spectrum of ${\cal H}_{N+1}$. The NRG eigenstates
so obtained are expected to faithfully describe
the spectrum of ${\cal H}_N$ on a scale of
$D_N = D_{\Lambda} \Lambda^{-(N-1)/2}$, corresponding
to the temperature $T_N \sim D_N$.
Thus, three distinct approximations are involved in
the NRG algorithm when applied to the Hamiltonian
of Eq.~(\ref{H-ep}):
(i) Discretization of the conduction band, controlled
    by the parameter $\Lambda > 1$;
(ii) A finite-size representation of the bare bosonic
     spectrum, controlled by the number $N_{\rm b}$
     of bare bosonic states kept (we use the states
     where $b^{\dagger}b = 0, \cdots, N_{\rm b}-1$);
(iii) Truncation of the Hilbert space at the conclusion
      of each NRG iteration, controlled by the number
      $N_{\rm s}$ of states retained.
Each of these three approximations can be systematically
improved by varying $\Lambda$, $N_{\rm b}$, and
$N_{\rm s}$. All data points presented in this paper
were obtained for $\Lambda = 2$, while the number
of states retained were either $N_{\rm b} = 1600$
and $N_{\rm s} = 4000$ or $N_{\rm b} = 3000$
and $N_{\rm s} = 8000$. Explicit values are
quoted in the relevant figure captions.

We emphasize that the total number of electrons,
or $Q$ in the notation of Ref.~\onlinecite{KWW80a},
is the only conserved quantity one can exploit in the
iterative diagonalization of the sequence of NRG
Hamiltonians for our problem. Since each additional site
can either be empty or occupied, it is straightforward
to keep track of the associated quantum number using the
algorithm detailed, e.g., in Ref.~\onlinecite{KWW80a}.

\begin{figure}[tb]
\centerline{
\includegraphics[width=80mm]{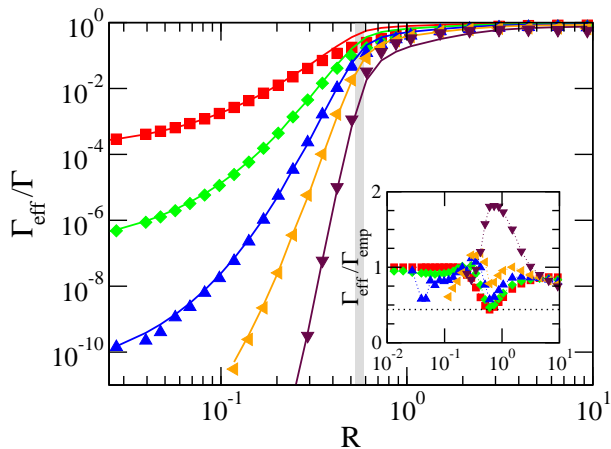}
}\vspace{0pt}
\caption{(Color online)
         The ratio $\Gamma_{\rm eff}/\Gamma$ vs
         $R = \Gamma/E_{\rm p}$, for
         $\omega_0/D = 0.0025$ and different strengths
         of the electron-phonon coupling $\lambda$.
         Here, $\lambda$ takes the values $3$ (red),
         $4$ (green), $5$ (blue), $6$ (orange), and $8$
         (maroon). Symbols depict the NRG data. Solid
         lines show the ratio $\Gamma_{\rm emp}/\Gamma$,
         obtained using the empirical formula
         $\Gamma_{\rm emp} = \Gamma\exp[-\lambda^2
         {\cal F}(R)]$ with ${\cal F}(R)$ given by the
         solid line in Fig.~\ref{Fig:Scaling}.
         The gray shaded area covers the regime where
         the condition $\Gamma_{\rm eff} = \omega_0$ is
         met for all values of $\lambda$ displayed.
         Inset: The ratio
         $\Gamma_{\rm eff}/\Gamma_{\rm emp}$ for
         all data points displayed in the main
         panel. The empirical formula and NRG data
         agree to within a factor of $2.3$ for
         all values of $\lambda$ and all ratios
         $\Gamma/E_{\rm p}$ displayed (the
         dotted line marks the value 0.44).
         NRG parameters: $\Lambda = 2$, $N_{\rm s}
         = 8000$, and $N_{\rm b} = 3000$.}
\label{Fig:G_eff}
\end{figure}

\subsection{Renormalized tunneling rate $\Gamma_{\rm eff}$}
\label{Sec:G_eff}

\begin{figure}[tb]
\centerline{
\includegraphics[width=80mm]{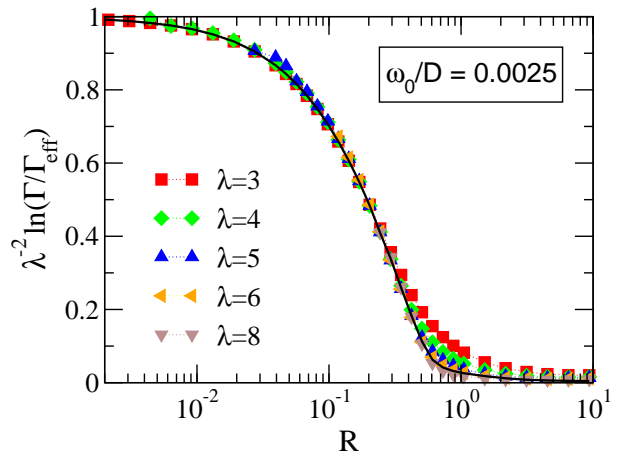}
}\vspace{0pt}
\caption{(Color online) The function
         $\lambda^{-2} \ln (\Gamma/\Gamma_{\rm eff})$
         vs $R = \Gamma/E_{\rm p}$, for
         $\omega_0/D = 0.0025$ and different strengths
         of the electron-phonon coupling $\lambda$.
         The solid line is an empirical curve that
         defines the function ${\cal F}(R)$, which
         is used in Fig.~\ref{Fig:G_eff} to compute
         $\Gamma_{\rm emp} =
         \Gamma\exp[-\lambda^2 {\cal F}(R)]$.
         All NRG parameters are the same as in
         Fig.~\ref{Fig:G_eff}.}
\label{Fig:Scaling}
\end{figure}

We begin our discussion with the renormalized tunneling
rate $\Gamma_{\rm eff}$, defined by Eqs.~(\ref{G_eff-def})
and (\ref{chi_c-def}) with $\chi_c$ evaluated for
$T \to 0$. Figure~\ref{Fig:G_eff} shows the dependence
of $\Gamma_{\rm eff}$ on $\Gamma$ for
$\omega_0/D = 0.0025$ and different strengths of the
electron-phonon coupling $\lambda$. As expected, the
ratio $\Gamma_{\rm eff}/\Gamma$ varies by orders of
magnitude upon going from small to large values of $\Gamma$
when $\lambda$ is large. For $E_{\rm p} \ll \Gamma$,
one essentially recovers the bare tunneling rate
$\Gamma$, while for $\Gamma \ll \omega_0$ there is
an exponential suppression of the tunneling rate
according to $\Gamma_{\rm eff}/\Gamma = e^{-\lambda^2}$.
Remarkably, the crossover between these two limits
follows an approximate scaling form, as demonstrated
in Fig.~\ref{Fig:Scaling}. Plotting
$\lambda^{-2} \ln (\Gamma/\Gamma_{\rm eff})$
as a function of $R = \Gamma/E_{\rm p}$
for the different values of $\lambda$, all data
points approximately collapse onto a single curve.
The collapse is particularly good in the range
$R \alt 0.2$, and gradually degrades for larger
values of $R$. Note that $E_{\rm p}/D$ equals $0.16$
for $\lambda = 8$, hence some values of $\Gamma$
become of order the bandwidth for $R \sim 10$.

The quality of the approximate scaling form (which,
as we show below, is not an exact scaling function)
can be appreciated by extracting an empirical function
${\cal F}(R)$ such that ${\cal F}(R) \approx
\lambda^{-2} \ln (\Gamma/\Gamma_{\rm eff})$, and
comparing the calculated values of $\Gamma_{\rm eff}$
to the empirical formula
\begin{equation}
\Gamma_{\rm emp} = \Gamma\exp[-\lambda^2 {\cal F}(R)] .
\label{G_emp}
\end{equation}
The empirical formula, depicted by the solid lines in
Fig.~\ref{Fig:G_eff}, well agrees with the calculated
values of $\Gamma_{\rm eff}$ over many orders of
magnitude. Deviations are confined to within a factor
of $2.3$ (see inset of Fig.~\ref{Fig:G_eff}), which is
quite remarkable considering the enormous variation in
$\Gamma_{\rm eff}$ as a function of both $\lambda$ and
$\Gamma$. As for the function ${\cal F}(R)$, we define
it by the solid line in Fig.~\ref{Fig:Scaling}.
Obviously, there is some arbitrariness in the way
${\cal F}(R)$ is fixed, particularly for $R \agt 0.2$
where scaling degrades. Indeed, it is in this parameter
regime that the deviations between $\Gamma_{\rm emp}$ and
$\Gamma_{\rm eff}$ are typically the largest. Still,
Eq.~(\ref{G_emp}) provides a useful formula for the
renormalized tunneling rate, successfully interpolating
between the extreme adiabatic and antiadiabatic
limits. Finally, we note that the condition
$\Gamma_{\rm eff} = \omega_0$ is met for $R$ in
the range $0.53 < R < 0.59$ (gray shaded area in
Fig.~\ref{Fig:G_eff}) for all values of $\lambda$
displayed, hence the two scales remain well
separated up to $R \sim 0.4$.

\subsection{Extended antiadiabatic limit: Mapping
            onto the IRLM}
\label{Sec:Mapping-parameters}

\begin{figure}[tb]
\centerline{
\includegraphics[width=75mm]{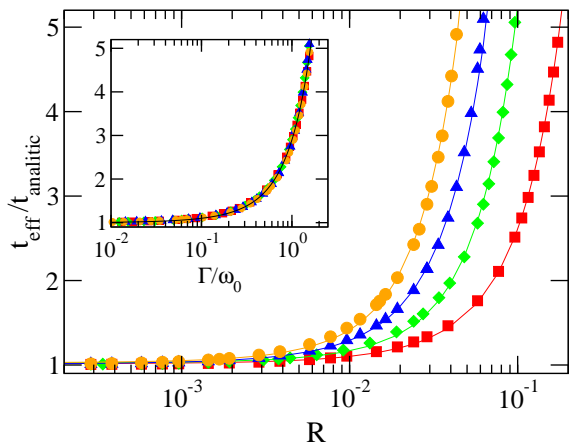}
}\vspace{0pt}
\caption{(Color online)
         The tunneling amplitude $t_{\rm eff}$
         that enters the effective IRLM at energy
         $D_{\rm eff}/\omega_0 = 0.146$, plotted
         vs $R = \Gamma/E_{\rm p}$. Here
         $\omega_0/D = 0.0025$, while $D_{\rm eff}$
         equals $D_{N = 22}$ with $\Lambda = 2$.
         The values of $\lambda$ are $3$ (red), $4$
         (green), $5$ (blue), and $6$ (orange). The
         remaining NRG parameters are $N_{\rm s} = 8000$
         and $N_{\rm b} = 3000$. For clarity,
         $t_{\rm eff}$ was normalized by
         $t_{\rm analytic} = t\, e^{-\lambda^2/2}$,
         which is the effective tunneling amplitude
         obtained for $\Gamma \ll \omega_0$.
         Inset: A scaling plot of
         $t_{\rm eff}/t_{\rm analytic}$ vs
         $\Gamma/\omega_0$. The solid line shows
         the parabola $f(x) = 1 + 0.9 x + x^2$.}
\label{Fig:t_eff}
\end{figure}

\begin{figure}[tb]
\centerline{
\includegraphics[width=75mm]{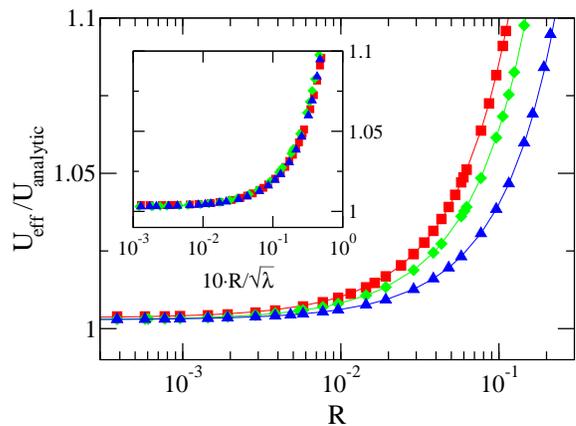}
}\vspace{0pt}
\caption{(Color online)
         The local Coulomb repulsion $U_{\rm eff}$
         that enters the effective IRLM at energy
         $D_{\rm eff}/\omega_0 = 0.146$, plotted
         vs $R = \Gamma/E_{\rm p}$. The values of
         $\lambda$ are $5$ (red), $10$ (green), and
         $20$ (blue). All other parameters are the
         same as in Fig.~\ref{Fig:t_eff}. For
         clarity, $U_{\rm eff}$ was normalized by
         $U_{\rm analytic} = 2 R/(\pi \rho_0)$,
         which is the effective Coulomb repulsion for
         $\Gamma \ll \omega_0$. Inset: A scaling plot
         of $U_{\rm eff}/U_{\rm analytic}$ versus
         $R/\sqrt{\lambda}$.}
\label{Fig:U_eff}
\end{figure}

Next we focus on the resonance condition
$\Delta \epsilon_d = 0$ and examine in greater detail
the regime where $\Gamma_{\rm eff} \ll \omega_0$. For
the case of interest where $1 \ll \lambda$, this condition
corresponds to $\Gamma \ll E_{\rm p}$, which
encompasses both the antiadiabatic limit $\Gamma \ll
\omega_0$ and an extended region where $\Gamma$ exceeds
$\omega_0$ and yet $\Gamma_{\rm eff} \ll \omega_0$. As
discussed in Sec.~\ref{Sec:Extended-AAL}, we anticipate
for such couplings that the system is described at
energies below $\omega_0$ by an effective IRLM with
the tunneling amplitude $t_{\rm eff}$ and the
local Coulomb repulsion $U_{\rm eff}$. Using the NRG
level flow, we have confirmed this physical
picture. In the intermediate
energy regime $\Gamma_{\rm eff} \ll D_N < \omega_0$,
the finite-size spectra consistently reduced to that
of a weakly coupled IRLM, proving the validity of the
extended antiadiabatic regime. The coupling constants
that enter the effective Hamiltonian can be read off
from the NRG spectra using the procedure outlined in
Appendix~\ref{app:IRLM-from-NRG}. Our results are
summarized in Figs.~\ref{Fig:t_eff} and \ref{Fig:U_eff}.

Figure~\ref{Fig:t_eff} displays the effective
tunneling amplitude $t_{\rm eff}$ at energy
$D_{\rm eff}/\omega_0 = 0.146$ for several values
of $\lambda = 2$, $4$, $5$, and $6$.
For clarity, all curves have been normalized by
$t_{\rm analytic} = t \, e^{-\lambda^2/2}$, which
accounts for the main Gaussian dependence of
$t_{\rm eff}$ on $\lambda$. As can be seen in the inset,
all data points for $t_{\rm eff}/t_{\rm analytic}$
collapse onto a single curve when plotted versus
$\Gamma/\omega_0$, at least for values of $\Gamma$
up to a few times $\omega_0$. Thus, the effective
tunneling amplitude acquires the empirical scaling form
\begin{equation}
t_{\rm eff} = t\cdot
              \exp\! \left [
                           -\frac{\lambda^2}{2}
                   \right ]
              f(\Gamma/\omega_0) ,
\label{t_eff-empirical}
\end{equation}
where $f(x)$ is well fitted by the parabola
$f(x) \approx 1 + 0.9 x + x^2$ (depicted by the full
line in the inset of Fig.~\ref{Fig:t_eff}). Three
points are noteworthy. First, $t_{\rm eff}$ grows
quite rapidly with $\Gamma/\omega_0$, increasing
by a factor of $5$ in the limited range covered by
the inset of Fig.~\ref{Fig:t_eff}. Second, given the
apparent scaling of $t_{\rm eff}/t_{\rm analytic}$
with $\Gamma/\omega_0$, it is evident that the ratio
$\Gamma_{\rm eff}/\Gamma$ cannot be an exclusive
function of $\Gamma/E_{\rm p}$ as suggested
by the scaling plot of Fig.~\ref{Fig:Scaling}. Lastly,
we cannot reliably extract $t_{\rm eff}$ for larger
values of $\Gamma/\omega_0$ since $t_{\rm eff}$ and
$U_{\rm eff}$ no longer represent well-separated
energy scales (for technical details, see
Appendix~\ref{app:IRLM-from-NRG}).

The effective Coulomb repulsion $U_{\rm eff}$
at energy $D_{\rm eff}/\omega_0 = 0.146$
is shown in turn in Fig.~\ref{Fig:U_eff}, after
division by its asymptotic weak-tunneling form
$U_{\rm analytic} = 2 R/(\pi \rho_0)$ [see
Eq.~(\ref{U_eff-weak-t}) with $1 \ll \lambda$].
Note that in contrast to $t_{\rm eff}$, which becomes
numerically inaccessible for $8 \alt \lambda$, the
effective Coulomb repulsion $U_{\rm eff}$ can be
accurately computed for values of $\lambda$ well above
$10$. Surprisingly, the weak-tunneling expression
that was derived strictly speaking for
$\Gamma \ll \omega_0$ remains quite accurate (to
within 10\%) even for $\Gamma/\omega_0$ as large as
$100$ when $\lambda = 20$. Hence, the main source
of $\Gamma$ dependence stems from $U_{\rm analytic}$
which scales as $\Gamma/E_{\rm p}$.
Similar to $t_{\rm eff}/t_{\rm analytic}$ also
$U_{\rm eff}/U_{\rm analytic}$ appears to follow
an approximate scaling form, this time with the
scaling variable $\Gamma/(\lambda^{2.5} \omega_0)$
(see inset of Fig.~\ref{Fig:U_eff}). However, the
quality of the data collapse and the variation in
$U_{\rm eff}/U_{\rm analytic}$ are far more
restricted than for $t_{\rm eff}/t_{\rm analytic}$.

From the discussion above it is clear that the
regimes $\Gamma \ll \omega_0$ and
$\omega_0 < \Gamma \ll E_{\rm p}$ share the same
qualitative physics, both being described by the same
IRLM at energies below $\omega_0$. The distinction
between the two regimes is mainly quantitative, as
encoded in the effective model parameters
$t_{\rm eff}$ and $U_{\rm eff}$. The rather rapid
departure of $t_{\rm eff}$ from its asymptotic
weak-tunneling form $t\, e^{-\lambda^2/2}$ reflects
its extreme sensitivity to even small deformations
of the polaronic mode.

\subsection{Charging of the level}
\label{Sec:Charging}

\begin{figure}[tb]
\centerline{
\includegraphics[width=75mm]{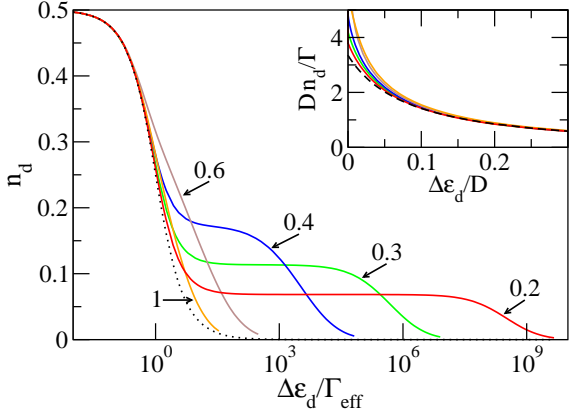}
}\vspace{0pt}
\caption{(Color online)
         The level occupancy $n_d$ vs
         $\Delta \epsilon^{}_d = \epsilon^{}_d
         - \epsilon_d^{\ast} > 0$, for
         $\omega_0/D = 0.0025$, $T \to 0$,
         $\lambda = 6$, and different values of
         $R = \Gamma/E_{\rm p}$ (indicated
         by arrows). The level occupancy in the
         complementary regime $\Delta \epsilon^{}_d < 0$
         is obtained from the symmetry relation of
         Eq.~(\ref{n_d-symmetry}). With increasing
         $\Gamma \geq E_{\rm p}$, charging
         of the level approaches the conventional
         noninteracting curve, depicted by the dotted
         line. With decreasing $\Gamma < E_{\rm p}$,
         the noninteracting shape rapidly deforms
         into a double-step structure governed by
         distinct mechanisms for
         $\Delta \epsilon^{}_d \alt \Gamma_{\rm eff}$
         and $\Gamma_{\rm eff} \ll \Delta \epsilon^{}_d$.
         For $\Delta \epsilon^{}_d \alt \Gamma_{\rm eff}$,
         one recovers a strongly renormalized
         noninteracting form with
         $\Gamma \to \Gamma_{\rm eff}$.
         For $\Gamma_{\rm eff} \ll \Delta \epsilon^{}_d$,
         charging is well described by a simple
         perturbative expansion in $t$. This is
         demonstrated in the inset, where each of
         the occupancies of the main panel is
         compared to the perturbative expression of
         Eq.~(\ref{n_d-perturbative}) (dashed line).
         NRG parameters: $\Lambda = 2$,
         $N_{\rm s} = 4000$, and $N_{\rm b} = 1600$.}
\label{Fig:charging}
\end{figure}

Up until now, our discussion was restricted to
$\Delta \epsilon_d = 0$. Next we consider nonzero
detuning and examine the charging properties of the
$d$ level. Besides being of interest on its own right,
the charge of the level is intimately related at $T = 0$
to the conductance of the molecular bridge depicted
schematically in Fig.~\ref{Fig:tunnel-junction}.
We address the latter setup in detail in
Sec.~\ref{Sec:conductance}.

Figure~\ref{Fig:charging} shows the level occupancy
$n_d$ versus $\Delta \epsilon_d >0$, for $T \to 0$ and
$\lambda = 6$. The level occupancy in the complementary
regime $\Delta \epsilon^{}_d < 0$ is obtained from
the symmetry relation of Eq.~(\ref{n_d-symmetry}).
With increasing $\Gamma \geq E_{\rm p}$,
charging of the level approaches the conventional
noninteracting curve, depicted by the dotted
line. With decreasing $\Gamma < E_{\rm p}$, the
noninteracting shape rapidly deforms into a charging
curve governed by two distinct mechanisms:
(i) a strongly renormalized noninteracting form with
   $\Gamma \to \Gamma_{\rm eff}$, applicable up to
   $\Delta \epsilon^{}_d \sim \Gamma_{\rm eff}$,
and
(ii) a simple perturbative expansion in $t$,
     applicable for $\Gamma_{\rm eff} \ll
     \Delta \epsilon^{}_d$.
The latter mechanism is demonstrated in the inset,
where each of the curves of the main panel is
compared to the perturbative expression of
Eq.~(\ref{n_d-perturbative}) (dashed line).

The combination of these two mechanisms gives
rise to a distinctive shoulder in $n_d$, which
interpolates between the limits where
$\Delta \epsilon^{}_d \sim \Gamma_{\rm eff}$ and
$\Delta \epsilon^{}_d \sim \Gamma$. The height of
the shoulder decreases with decreasing $\Gamma$,
approaching $\Gamma/(\pi E_{\rm p})$ when
$\Gamma \ll E_{\rm p}$. This result can
be understood from the fact that the summation
over $n$ in Eq.~(\ref{n_d-perturbative}) samples
mainly the regime where $n \sim \lambda^2$ when
$\lambda$ is large, hence the denominator
$\Delta \epsilon_d + n \omega_0$ is effectively
replaced with $\Delta \epsilon_d + \lambda^2 \omega_0
\approx E_{\rm p}$.  This argumentation breaks
down as $\Delta \epsilon_d$ approaches
$\Gamma_{\rm eff}$, when higher order terms
become exceedingly more important. Another
effect of decreasing $\Gamma$ is the opening
of an exponential separation between $\Gamma$ and
$\Gamma_{\rm eff}$, which sets the lateral extent
of the shoulder when plotted versus
$\Delta \epsilon_d^{}/\Gamma_{\rm eff}$.
For example, $\Gamma$ and $\Gamma_{\rm eff}$
are separated by one order of magnitude for
$R = 0.6$, which is insufficient for a
fully developed shoulder to be seen.

\subsection{Phononic distribution}
\label{Sec:P(n)}

\begin{figure}[tb]
\centerline{
\includegraphics[width=75mm]{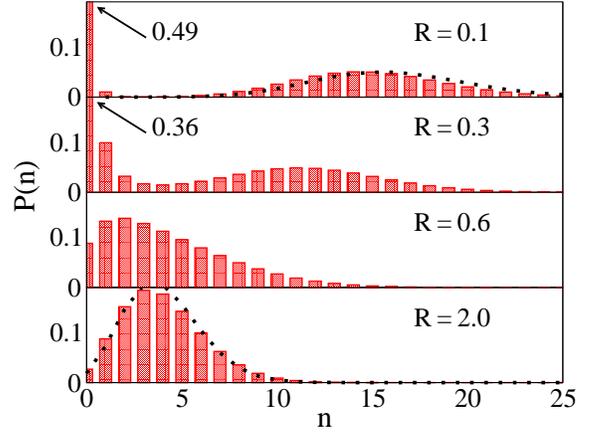}
}\vspace{0pt}
\caption{(Color online)
         The phonon distribution function
         $P(n) = \left < |n \rangle \langle n | \right >$,
         for $\omega_0/D = 0.0025$, $\lambda = 4$,
         $N_0 = 0$, $T \to 0$, and different values
         of $R = \Gamma/E_{\rm p}$. The detuning
         energy $\Delta \epsilon^{}_d$ is set to zero
         such that the electronic level is half filled.
         For $R \ll 1$, exemplified by $R = 0.1$ in
         the upper panel, the distribution function
         $P(n)$ approaches the double-peak structure
         of Eq.~(\ref{P(n)-AAL}), depicted by the dotted
         line. In the opposite limit where $R$ is large,
         exemplified by $R = 2$ in the lower panel, the
         distribution function $P(n)$ approaches the
         single-peak structure of Eq.~(\ref{P(n)-AL}),
         displayed by the dotted line. NRG parameters:
         $\Lambda = 2$, $N_{\rm s} = 4000$, and
         $N_{\rm b} = 1600$.}
\label{Fig:Ph-dis}
\end{figure}

To gain direct information on the state of the phonon,
we next consider the phonon distribution function
$P(n)$, defined in Eq.~(\ref{P(n)-def}).
Figure~\ref{Fig:Ph-dis} depicts $P(n)$ for $T \to 0$
using $\lambda = 4$, $N_0 = 0$, and different values
of $R = \Gamma/E_{\rm p}$. The detuning energy
$\Delta \epsilon_d^{}$ (which itself depends on $N_0$
through $\epsilon_d^{\ast}$) is set to zero, corresponding
to resonance condition. For small values of $\Gamma$,
exemplified by $R = 0.1$ in the upper panel, the
distribution function $P(n)$ approaches the
double-peak structure of Eq.~(\ref{P(n)-AAL}),
which is plotted for comparison by the dotted line.
Although $\Gamma$ equals $1.6 \omega_0$ in this case,
the phonon distribution function remains well described
by the antiadiabatic result which applies, strictly
speaking, to $\Gamma \ll \omega_0$. Physically this
means that the degree of undressing of the polaron is
rather small for this particular value of $\Gamma$.
A qualitative change in the profile of $P(n)$ takes
place upon going from $R = 0.3$ to $R = 0.6$,
marking the progressive undressing of the polaron.
Finally for large $\Gamma$, exemplified by $R = 2$
in the lower panel, the distribution function $P(n)$
approaches the single-peak structure of
Eq.~(\ref{P(n)-AL}), which is plotted for
comparison by the dotted line. As discussed in
Sec.~\ref{Sec:P(n)-analytic}, the phonon is too
slow to track the individual tunneling events in this
limit, hence it samples only the average displacement
$\lambda \langle \hat{n}_d \rangle = \lambda/2$.

A useful perspective on the phononic state is provided
by varying the reference charge $N_0$ of the electronic
level. As discussed in Sec.~\ref{Sec:model}, $N_0$ can
be eliminated from the Hamiltonian of Eq.~(\ref{H-ep})
by simply shifting the bosonic mode according to
$\hat{B} = b - \lambda N_0$. If the level energy
$\epsilon^{}_d$ is simultaneously adjusted such
that $\Delta \epsilon^{}_d = \epsilon^{}_d -
\lambda^2 \omega_0 ( 1 - 2N_0 )$
is held fixed, then the variation of $N_0$ has no
effect on the low-energy spectrum of the Hamiltonian.
Nevertheless, the low-energy state of the system does
vary with $N_0$, as can be seen from the case
where the $\hat{B}$ boson occupies a pure
coherent state. If $\hat{B}$ resides in the coherent
state $|z \rangle$, then the $b$ phonon occupies the
shifted coherent state $|z + \lambda N_0\rangle$. In
other terms, one can continuously vary the coherent
state that $b$ occupies by tuning $N_0$ while holding
$\Delta \epsilon^{}_d$ fixed. This strategy can be used
not only to expose a pure coherent state, but also to
diagnose a superposition involving a small number
of coherent states of the type often used in
variational calculations.

\begin{figure}[tb]
\centerline{
\includegraphics[width=75mm]{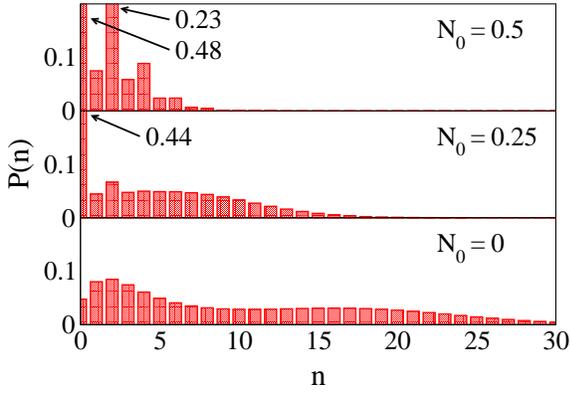}
}\vspace{0pt}
\caption{(Color online)
         The phonon distribution function
         $P(n) = \left < |n \rangle \langle n | \right >$,
         for $\omega_0/D = 0.0025$, $\lambda = 6$,
         $R = \Gamma/E_{\rm p} = 0.5$, $T \to 0$, and
         different values of $N_0$.  The detuning
         energy $\Delta \epsilon^{}_d$ is held fixed
         at zero for all values of $N_0$, such that
         the electronic level is half filled.
         NRG parameters: $\Lambda = 2$,
         $N_{\rm s} = 4000$, and $N_{\rm b} = 1600$.}
\label{Fig:Ph-dis-N0}
\end{figure}

Figure~\ref{Fig:Ph-dis-N0} displays the evolution
of the phonon distribution function with increasing
$N_0$, for $\lambda = 6$ and fixed
$\Delta \epsilon_d^{} = 0$. The ratio
$\Gamma/E_{\rm p} = 0.5$ is tuned to the middle
of the crossover regime, such that the system
is well removed from the extreme adiabatic and
antiadiabatic limits. As can be seen, $P(n)$
evolves in a rather complicated manner upon
going from $N_0 = 0$ to $N_0 = 0.5$. Initially,
there are two rather broad and smooth humps for
$N_0 = 0$. Upon increasing $N_0$, the phonon
distribution function narrows considerably,
developing multiple sharp peaks by the time
$N_0 = 0.5$. Indeed, $3$ clear maxima are
visible at $n = 0$, $2$, and $4$ when
$N_0 = 0.5$, indicating that at least three
independent coherent states significantly
contribute to the low-energy state of the
system. We have not succeeded in reproducing
this multiple-peak structure for $N_0 = 0.5$
(let alone the entire collection of
distributions for the different values of
$N_0$) by employing a simple superposition of
just a few coherent states. Given the sharpness
of the peaks found at even values of $n$, it
is clear that the low-energy state of the system
can not be well represented using just one or
two coherent states. This sets a stringent
constraint on variational treatments of the
crossover regime when $\lambda$ is large.

\section{Conductance of a molecular bridge}
\label{Sec:conductance}

\begin{figure}[tb]
\centerline{
\includegraphics[width=75mm]{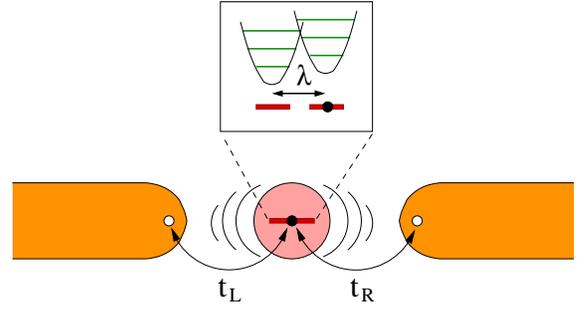}
}\vspace{0pt}
\caption{(Color online)
         Schematic description of a molecular
         bridge, where a single molecule is
         trapped between two leads.}
\label{Fig:tunnel-junction}
\end{figure}

Our discussion thus far has focused on thermodynamic
quantities, which are difficult to measure for a
single molecule. Of particular interest are transport
properties of molecular devices of the type displayed
schematically in Fig.~\ref{Fig:tunnel-junction}, where
a single molecule is trapped between two leads.
Continuing with spinless fermions, we model such a
molecular bridge by the Hamiltonian
\begin{align}
{\cal H} &=
         \sum_{k, \alpha}
               \epsilon_k c^{\dagger}_{k \alpha}
                          c^{}_{k \alpha}
         + \sum_{\alpha}
               \frac{t_{\alpha}}{\sqrt{N}}
               \sum_{k}
                    \left \{
                             d^{\dagger} c^{}_{k \alpha}
                             + {\rm H.c.}
                    \right \}
\nonumber \\
        &+ \epsilon^{}_d d^{\dagger} d
         + \omega_{0} b^{\dagger} b
         + \lambda \omega_0\!
           \left (
                   d^{\dagger} d - N_0
           \right )\!
           \left (
                   b^{\dagger} + b
           \right ) ,
\label{H-bridge}
\end{align}
where $\alpha = L, R$ labels the left/right lead.
Here, $c^{\dagger}_{k \alpha}$ creates a conduction
electron with momentum $k$ in lead $\alpha$, while
$t_{\alpha}$ is the tunneling matrix element between
the electronic level and the Wannier state closest to
the molecule in lead $\alpha$. For convenience we
take the number of lattice sites $N$ to be the same
in both leads, although this condition can easily be
relaxed.

In equilibrium, the two-lead Hamiltonian of
Eq.~(\ref{H-bridge}) is equivalent to the single-band
model of Eq.~(\ref{H-ep}). This stems from the fact
that the localized level couples solely to the
``bonding'' combination
\begin{equation}
c^{\dagger}_{+ k} =
             \frac{t_L}{\sqrt{t_L^2 + t_R^2}}\,
             c^{\dagger}_{k L} +
             \frac{t_R}{\sqrt{t_L^2 + t_R^2}}\,
             c^{\dagger}_{k R} ,
\end{equation}
while the ``anti-bonding'' combination
\begin{equation}
c^{\dagger}_{- k} =
             \frac{t_R}{\sqrt{t_L^2 + t_R^2}}\,
             c^{\dagger}_{k L} -
             \frac{t_L}{\sqrt{t_L^2 + t_R^2}}\,
             c^{\dagger}_{k R}
\end{equation}
is decoupled from the level. Inasmuch as
impurity-related quantities are concerned, one can
omit then the ``anti-bonding'' band altogether,
to be left with the single-band Hamiltonian of
Eq.~(\ref{H-ep}) where $t = \sqrt{t_L^2 + t_R^2}$.

\begin{figure}[tb]
\centerline{
\includegraphics[width=75mm]{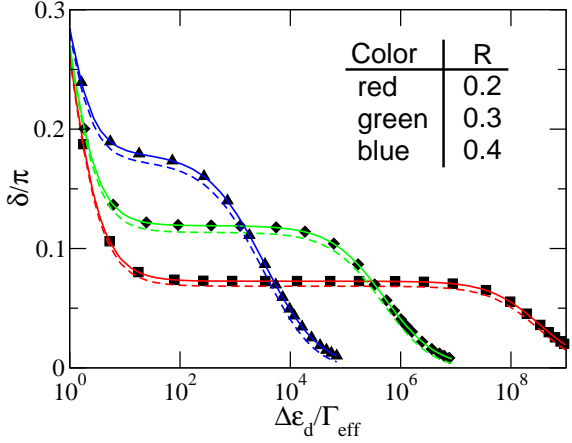}
}\vspace{0pt}
\caption{(Color online)
         The scattering phase shift $\delta$ vs
         the detuning energy $\Delta \epsilon^{}_d$,
         for $\omega_0/D = 0.0025$, $\lambda = 6$,
         $T \to 0$, and three intermediate values of
         $R = \Gamma/E_{\rm p}$. Here $\Gamma =
         \Gamma_L + \Gamma_R$ is the total tunneling
         rate from the level to the two leads. Solid
         lines depict the phase shift $\delta/\pi$,
         dashed lines show the dot occupancy $n_d$,
         and the symbols show the number of displaced
         electrons, $\Delta N$. In accordance with
         the Friedel-Langreth sum rule, $\delta/\pi$
         coincides with $\Delta N$, which deviates,
         however, from $n_d$.
         NRG parameters: $\Lambda = 2$,
         $N_{\rm s} = 4000$, and $N_{\rm b} = 1600$.}
\label{Fig:Phase-shift}
\end{figure}

The restriction to the ``bonding'' band is no longer
complete when a finite bias is applied between the
leads, since the nonequilibrium boundary condition
applies to the lead electrons rather than their
``bonding'' and ``anti-bonding'' combinations. Thus,
one can no longer settle with the single-band
Hamiltonian of Eq.~(\ref{H-ep}) for a biased junction.
Fortunately, this complication can be circumvented
in linear response, when the conductance can be
expressed in terms of equilibrium response functions
of the system. In particular, the zero-temperature
conductance, $G$, is determined by the scattering
phase shift of the ``bonding'' electrons, as follows
from the Fermi-liquid ground state of the system.
The zero-temperature conductance thus takes the
standard form
\begin{equation}
G = G_0 \sin^{2}(\delta) ,
\end{equation}
where
\begin{equation}
G_0 = \frac{e^2}{h}
      \frac{4 \Gamma_L \Gamma_R}
           {(\Gamma_L + \Gamma_R)^2}
\label{G_0}
\end{equation}
with $\Gamma_{\alpha} = \pi \rho_0 t_{\alpha}^2$ is
a geometric factor encoding the degree of asymmetry
in the coupling to the two leads and $\delta$ is the
scattering phase shift. Generalizing the Friedel-Langreth
sum rule~\cite{Langreth66} to the present setting,
the phase shift $\delta$ is given by
\begin{equation}
\delta = \pi \Delta N ,
\end{equation}
where $\Delta N$ is the number of displaced electrons.
The latter quantity comprises two contributions:
(i) occupancy of the level $n_d$, and
(ii) the change in occupancy of the ``bonding'' band
     inflicted by the coupling to the level.
In the wide-band limit only the former contribution
is left, resulting in $\delta = \pi n_d$. However,
as we show below, $\Delta N$ deviates from $n_d$
when the conduction-electron bandwidth is finite.

Figure~\ref{Fig:Phase-shift} shows the scattering phase
shift $\delta/\pi$ versus $\Delta \epsilon_d^{}$,
calculated directly from the NRG spectra using the standard
prescription~\cite{ALPC92} of Eq.~(\ref{delta-from-eta}).
For comparison, we also plot the level occupancy $n_d$
and the number of displaced electrons $\Delta N$,
which is readily computed in the NRG since the total
electronic occupancy of the system is a good quantum
number that is kept track of in the course of the
iterative procedure. The
electron-phonon coupling is set equal to $\lambda = 6$,
while $\Gamma = \Gamma_L + \Gamma_R$ is tuned to three
intermediate values of $R = \Gamma/E_{\rm p}$ where
$n_d$ shows a pronounced shoulder.

\begin{figure}[tb]
\centerline{
\includegraphics[width=75mm]{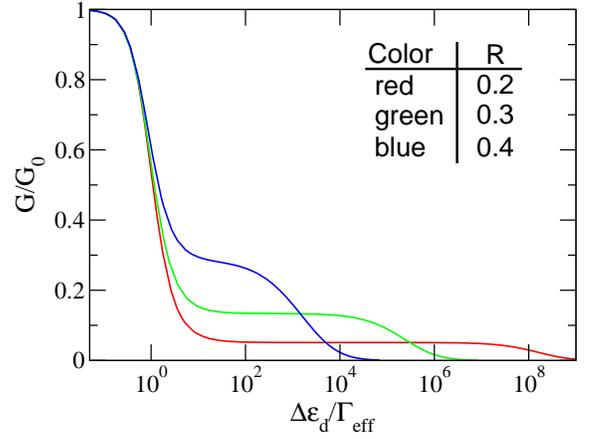}
}\vspace{0pt}
\caption{(Color online)
         The zero-temperature conductance $G$ vs
         the detuning energy $\Delta \epsilon^{}_d$,
         plotted for the same model and NRG parameters
         as in Fig.~\ref{Fig:Phase-shift}. Here $G_0$
         is a geometric factor specified in
         Eq.~(\ref{G_0}), encoding the asymmetry of
         the tunnel junction. The characteristic
         shoulder in $n_d$ translates to a similar
         shoulder in $G/G_0$, providing a distinct
         experimental fingerprint of phonon-assisted
         tunneling when the electron-phonon coupling
         is strong.}
\label{Fig:conductance}
\end{figure}

While $\delta/\pi$ and $\Delta N$ are indistinguishable
to within numerical accuracy, there is a small but
consistent deviation from $n_d$, particularly along
the characteristic shoulder in $n_d$. The difference
between $\Delta N$ and $n_d$ appears to somewhat
increase upon going from $R = 0.2$ to $0.4$, and
disappears for $\Delta \epsilon_d^{} \to 0$
(not shown). This latter behavior is to be expected
since $\delta/\pi$, $\Delta N$, and $n_d$ are all
pinned to $1/2$ by particle-hole symmetry when
$\Delta \epsilon_d^{} = 0$.

The corresponding zero-temperature conductance
is shown in turn in Fig.~\ref{Fig:conductance}
as a function of the detuning energy
$\Delta \epsilon_d^{}$. Note that the conductance
is an even function of $\Delta \epsilon_d^{}$,
as follows from the particle-hole transformation
of Eq.~(\ref{transformation}). Hence only the regime
$\Delta \epsilon_d^{} > 0$ is displayed. The
characteristic shoulder in $n_d$ translates to a
similar shoulder in
$G/G_0$, which is further pushed down in magnitude
due to the quadratic dependence on $\sin(\delta)$.
Experimentally, $\Delta \epsilon_d^{}$ can be tuned
using a suitable gate voltage, giving rise to
characteristic shoulders in the conductance as a
function of gate voltage. These shoulders, which
are quite unique in the context of tunneling through
confined nanostructures, provide a distinct experimental
fingerprint of phonon-assisted tunneling when the
electron-phonon coupling is strong.

\section{Summary}
\label{Sec:summary}

Focusing on strong electron-phonon interactions,
$1 \ll \lambda$, we presented a comprehensive study of
the crossover from the antiadiabatic to the adiabatic
regime of phonon-assisted tunneling in the framework
of a minimal model for molecular devices: a resonant
level coupled by displacement to a single localized
vibrational mode. Our main findings are as follows.
\begin{enumerate}
\item
   In contrast to common lore, the crossover from
   the polaronic physics of the antiadiabatic limit
   to the perturbative physics of the adiabatic regime
   is governed primarily by the polaronic shift
   $E_{\rm p} = \lambda^2 \omega_0$ rather than the
   phonon frequency $\omega_0$. In particular, the
   perturbative adiabatic limit is approached only as
   the bare hopping rate $\Gamma$ exceeds the polaronic
   shift, leaving an extended window of couplings where
   $\Gamma$ well exceeds the phonon frequency and yet
   the physics is basically that of the antiadiabatic
   regime.
\item
   Throughout the traditional and the extended
   antiadiabatic regimes, the effective low-energy
   Hamiltonian at energies below $\omega_0$ is the
   purely fermionic IRLM, which depends at resonance
   on two parameters only: $t_{\rm eff}$ and
   $U_{\rm eff}$. The effective tunneling amplitude
   $t_{\rm eff}$ obeys the empirical scaling form of
   Eq.~(\ref{t_eff-empirical}), at least up to values
   of $\Gamma$ several times larger than $\omega_0$,
   while $U_{\rm eff}$ is well approximated by
   Eq.~(\ref{U_eff}) for much of the extended
   antiadiabatic regime.
\item
   Although $\Gamma_{\rm eff}$ varies by many orders
   of magnitude as a function of $\lambda$ and
   $R = \Gamma/E_{\rm p}$, it is rather well described
   for all parameter regimes by the empirical formula
   of Eq.~(\ref{G_emp}), which depends on a single
   scaling function ${\cal F}(R)$. Our proposal for
   ${\cal F}(R)$ is given by the solid line in
   Fig.~\ref{Fig:Scaling}, although this choice can
   possibly by further optimized.
\item
   Charging properties are governed by two distinct
   mechanisms at the extended antiadiabatic and
   into the crossover regime. At small detuning,
   $\Delta \epsilon^{}_d \alt \Gamma_{\rm eff}$,
   the level occupancy $n_d$ follows a strongly
   renormalized noninteracting form with
   $\Gamma \to \Gamma_{\rm eff}$. By contrast,
   a simple perturbative expansion in $t$ applies
   for $\Gamma_{\rm eff} \ll \Delta \epsilon^{}_d$,
   giving rise to a characteristic shoulder in
   $n_d$ and in the low-temperature conductance of
   a molecular junction as a function of gate voltage.
\end{enumerate}
Our scenario for the crossover from the
antiadiabatic to the adiabatic regime is summarized
in Fig.~\ref{Fig:E-p}, where we have merged the
traditional and the extended antiadiabatic regimes
on the basis the two are qualitatively the same.

This study was devoted to thermodynamic
properties, for which a rather complete picture was
provided. However, the question of real-time dynamics
remains largely open. Particularly, does $\tau_{\rm p}
= 1/E_{\rm p}$ show up as a new time scale in the
dynamics? As could be anticipated, $\tau_{\rm p}$
plays no role deep in the adiabatic regime, when
$E_{\rm p} \ll \Gamma$. Indeed, as recently shown
for different scenarios of quench and driven
dynamics,~\cite{VSA12} only three time scales
are involved in this limit: the dwell time
$\tau_{\rm dwell} = 1/\Gamma$, the period of
oscillations $\tau_{\rm osc} = 2\pi/\omega_{\rm eff}$
with the softened frequency of
Eq.~(\ref{omega_eff}), and the phonon damping time
$\tau_{\rm damp} = \pi \Gamma^2/(\lambda^2 \omega_0^3)$,
extracted from the imaginary part of Eq.~(\ref{z_pm}).
Other perturbative~\cite{RS2009} and
numerical~\cite{Muehlbacher-rabani-2008,Thoss-2011,
Thoss-2012} studies of
real-time dynamics have focused primarily on the
build up of the current in a biased two-lead setting,
revealing rich behavior. Most notably, a significant
dependence on the initial state was reported in
Ref.~\onlinecite{Thoss-2012} up to times far
exceeding $\tau_{\rm dwell}$, suggesting the
existence of another, much longer time scale
whose origin is not quite clear. We note in
passing that the parameter set referred to as
adiabatic in the latter study corresponds in
our notation to $\lambda = 8$ and $R = 1/8$,
which actually falls in the extended antiadiabatic
regime. This may suggest the possible relevance at
long times of a strongly renormalized tunneling
rate akin to $\Gamma_{\rm eff}$.
Clearly the understanding of real-time dynamics
in molecular devices is a timely and challenging
task that deserves further investigation.

\section*{Acknowledgments}

We are grateful to E. Lebanon for stimulating our
interest in the problem. This research was supported
by the Israel Science Foundation through Grant
No. 1524/07 and by the German-Israeli Foundation
through Grant No. 1035-36.14.

\appendix

\section{Schrieffer-Wolff-type transformation for
         $\Gamma \ll \omega_0$}
\label{app:SWT}

In this Appendix, we describe the Schrieffer-Wolff-type
transformation that maps the Hamiltonian of
Eq.~(\ref{H-LF}) onto the IRLM Hamiltonian of
Eq.~(\ref{H-eff}), with the coupling constants
$t_{\rm eff}$ and $U_{\rm eff}$ specified in
Eqs.~(\ref{t_eff}) and (\ref{U_eff}), respectively.
As emphasized in the main text, the mapping applies to
$\Gamma \ll \omega_0$, when a single-step elimination
of all excitation energies exceeding $\omega_0$ is
sufficient. The mapping is further restricted to
temperatures below $\omega_0$, when the (transformed)
phonon is frozen in its unperturbed ground-state
configuration $b^{\dagger} b = 0$.

Following Schrieffer and Wolff,~\cite{SW66} we seek
a canonical transformation
\begin{equation}
{\cal H}_{\rm S} = e^S {\cal H}' e^{-S}
\end{equation}
with a suitable anti-Hermitian operator $S$ such that
the low-energy subspace is decoupled to order $t^2$
from all excited phononic states and all
single-particle conduction-electron excitations
with energy $|\epsilon_k| > \omega_0$ (whether
particle or hole). To this end, we
introduce two complementary projection operators,
$P$ and $Q = 1 - P$, where $P$ projects onto the
low-energy subspace with $b^{\dagger} b = 0$ and
no single-particle conduction-electron excitations
whose energy exceeds $\omega_0$. In other terms,
all conduction-electron modes with energy
$\epsilon_k < -\omega_0$ ($\omega_0 < \epsilon_k$)
are strictly occupied (unoccupied) within the
subspace defined by $P$. The Hamiltonian
${\cal H}'$ is then divided into an ``unperturbed''
part ${\cal H}'_0$ and a ``perturbation''
${\cal H}'_1$, where
\begin{equation}
{\cal H}'_0 = P {\cal H}' P
            + Q {\cal H}' Q
\end{equation}
and
\begin{equation}
{\cal H}'_1 = P {\cal H}' Q
            + Q {\cal H}' P .
\end{equation}
Explicitly, ${\cal H}'_0$ takes the form
\begin{align}
{\cal H}'_0 &= {\cal H}_0
            + \frac{t}{\sqrt{N}} e^{-\lambda^2/2}
              \sum_{k}^{\rm restric}
                   \left \{
                            d^{\dagger} c^{}_{k}
                            + {\rm H.c.}
                   \right \}
              |0 \rangle \langle 0| P
\nonumber \\
           &+ \frac{t}{\sqrt{N}}
              Q \sum_{k}
                     \left \{
                              e^{\lambda (b^{\dagger} - b)}
                              d^{\dagger} c^{}_{k}
                              + {\rm H.c.}
                     \right \} Q
\label{H'_0-explicit}
\end{align}
with
\begin{equation}
{\cal H}_0 = \sum_{k} \epsilon_k
                      c^{\dagger}_{k} c^{}_{k}
             + \omega_0 b^{\dagger} b ,
\end{equation}
while ${\cal H}'_1$ is given by
${\cal H}'_1 = \tilde{\cal H}'_1 P + {\rm H.c.}$ with
\begin{align}
\tilde{\cal H}'_1 &=
    \frac{t}{\sqrt{N}}
    \sum_{n > 0}
    \sum_k
          \left [
                  U^{-}_n c^{\dagger}_k d
                  + U^{+}_n d^{\dagger} c_k
          \right ]
          |n \rangle \langle 0|
\nonumber\\
  &+
    \frac{t}{\sqrt{N}}
    e^{-\lambda^2/2}
    \left [
            \sum_{k}^{\rm elec}
                 c^{\dagger}_k d
            + \sum_{q}^{\rm hole}
                   d^{\dagger} c_q
    \right ]
    |0 \rangle \langle 0|
\label{H'_1-explicit}
\end{align}
and
\begin{equation}
U^{\pm}_n = e^{-\lambda^2/2}
            \frac{(\pm \lambda)^n}{\sqrt{n!}} .
\end{equation}
Here the symbols $\sum_{k}^{\rm elec}$ and
$\sum_{q}^{\rm hole}$ come to indicate that the
summations over $k$ and $q$ are restricted to
momenta such that $\epsilon_k > \omega_0$ and
$\epsilon_q < -\omega_0$, respectively.

Using the formal expansion
\begin{equation}
{\cal H}_{\rm S} = {\cal H}'_0 + {\cal H}'_1
                 + [S, {\cal H}'_0] + [S, {\cal H}'_1]
                 + \frac{1}{2}
                   [ S, [S, {\cal H}'_0]] + \ldots
\label{Expansion-in-S}
\end{equation}
and anticipating that $S$ is proportional to $t$
at leading order (as will shortly be seen), one can
group the different terms in Eq.~(\ref{Expansion-in-S})
according to powers in $t$. The requirement that no
coupling is left to order $t^2$ between the excited
and low-energy subspaces is satisfied by demanding
that
\begin{equation}
{\cal H}'_1 + [S, {\cal H}'_0] = {\cal O}(t^3) ,
\label{Condition-on-S}
\end{equation}
resulting in
\begin{equation}
{\cal H}_{\rm S} = {\cal H}'_0
                 + \frac{1}{2}
                   [ S, {\cal H}'_1 ]
                 + {\cal O}(t^3) .
\label{H-2nd-order}
\end{equation}

Equation~(\ref{Condition-on-S}) has the formal
solution
\begin{equation}
S = -\frac{1}{{\cal L}'_0} {\cal H}'_1
  + {\cal O}(t^3) ,
\label{formal-solution-for-S}
\end{equation}
where ${\cal L}'_0$ is the Liouville operator defined
by ${\cal L}'_0 \hat{O} = [ \hat{O}, {\cal H}'_0]$.
Since ${\cal H}'_1$ is proportional to $t$, it is
clear that $S$ has a leading linear dependence on $t$,
as we have assumed. The anti-Hermitian operator $S$ does
contain, however, additional higher order terms in $t$,
which stem from the fact that ${\cal H}'_0$ (and thus
${\cal L}'_0$) includes components linear in $t$ [see
Eq.~(\ref{H'_0-explicit})]. Denoting the linear-order
component of $S$ by $S^{(1)}$, the latter is computed by
substituting ${\cal H}'_0 \to {\cal H}_0$, corresponding
to setting ${\cal L}'_0 \to {\cal L}_0$ with
${\cal L}_0 \hat{O} = [ \hat{O}, {\cal H}_0]$ in
Eq.~(\ref{formal-solution-for-S}). Using ${\cal H}'_1$
of Eq.~(\ref{H'_1-explicit}), this yields the explicit
expression
\begin{equation}
S^{(1)} = \tilde{S} P - {\rm H.c.}
\label{S-explicit-1}
\end{equation}
with
\begin{align}
\tilde{S} &=
    \frac{t}{\sqrt{N}}
    \sum_{n > 0}
    \sum_k
          \left [
                  c^{\dagger}_k d
                  \frac{U^{-}_n}
                       {\epsilon_k + n \omega_0}
                  + d^{\dagger} c_k
                  \frac{U^{+}_n}
                       {n \omega_0 - \epsilon_k}
          \right ]
          |n \rangle \langle 0|
\nonumber\\
  &+
    \frac{t}{\sqrt{N}}
    e^{-\lambda^2/2}
    \left [
            \sum_{k}^{\rm elec}
                 \frac{1}{\epsilon_k}
                 c^{\dagger}_k d
            - \sum_{q}^{\rm hole}
                   \frac{1}{\epsilon_q}
                   d^{\dagger} c_q
    \right ]
    |0 \rangle \langle 0| .
\label{S-explicit-2}
\end{align}

To obtain ${\cal H}_{\rm S}$ up to second order
in $t$, it suffices to replace $S$ in
Eq.~(\ref{H-2nd-order}) with $S^{(1)}$ of
Eqs.~(\ref{S-explicit-1}) and (\ref{S-explicit-2}).
Carrying out the commutator in Eq.~(\ref{H-2nd-order}),
projecting the result onto the $b^{\dagger} b = 0$
subspace, and restricting the band to
$|\epsilon_k| < \omega_0$, one arrives to order
$t^2$ at the effective low-energy Hamiltonian of
Eq.~(\ref{H-eff}) with the coupling constants
specified in Eqs.~(\ref{t_eff}) and (\ref{U_eff}).

Two comments should be made about the derivation of
the effective Hamiltonian of Eq.~(\ref{H-eff}). First,
we have explicitly assumed a particle-hole symmetric
band. Second, we have neglected $\epsilon_k$ in the
denominators on the first line of
Eq.~(\ref{S-explicit-2}), on the premise that
$|\epsilon_k|$ is small as compared to the new
ultraviolet cutoff energy $\omega_0$.

\section{Extracting the couplings of the effective IRLM
         Hamiltonian}
\label{app:IRLM-from-NRG}

As discussed in the main text, the Hamiltonian
of Eq.~(\ref{H-ep}) with $1 \ll \lambda$ and
$\Gamma \ll E_{\rm p}$ can be described at energies
below $\omega_0$ by an effective IRLM. For
$\Delta \epsilon^{}_d = 0$, the case considered
hereafter, particle-hole symmetry restricts the
number model parameters in the IRLM to two:
the tunneling amplitude $t_{\rm eff}$ and the
local Coulomb repulsion $U_{\rm eff}$.
In this Appendix, we describe in detail how these
model parameters can be extracted using the NRG.
The analysis relies on certain characteristics,
exposed below, of the finite-size spectrum of
the IRLM near the free-impurity fixed point,
$t_{\rm eff} = U_{\rm eff} = 0$.

\begin{table}[tb]
\begin{tabular}{c|c|c}
        \hline
        Energy level  & Quantum numbers
                      & Hopping matrix \\
        ($\Lambda = 2$) & $(Q_d ,Q_c)$ & element to \\
        \hline
        \multirow{2}{0.2cm}[-0.05cm]{0}
                                & $(-1,0)$ & --- \\
                                & $(+1,0)$ & --- \\
        \hline
        \multirow{4}{1cm}[-0.1cm]{0.4916}
                                & $(-1,-2)$ & --- \\
                                & $(+1,-2)$ & --- \\
                                & $(-1,+2)$ & --- \\
                                & $(+1,+2)$ & --- \\
        \hline
        \multirow{2}{0.8cm}[-0.05cm]{0.9832}
                                & $(-1,0)$ & --- \\
                                & $(+1,0)$ & --- \\ \hline
\end{tabular}
\caption{Finite-size NRG spectrum at the noninteracting
         free-impurity fixed point $t = U = 0$, for
         $\Lambda = 2$ and odd iteration number $N$.
         The left column specifies the dimensionless
         NRG energies. The corresponding quantum
         numbers, $Q_d$ and $Q_c$, are listed in the
         central column, while the right-hand-side
         column indicates degenerate eigenstates that
         are connect to the designated state with a
         nonzero hopping matrix element once a finite
         $t$ is switched on.}
\label{Table-odd}
\end{table}

In our analysis we shall use the IRLM in its
following representation
\begin{eqnarray}
{\cal H} &=&
             \sum_{k} \epsilon_k
                     c^{\dagger}_{k} c^{}_{k}
             + \frac{t}{\sqrt{N_k}}
                    \sum_{k}
                        \left \{
                                 d^{\dagger} c^{}_{k}
                                 + {\rm H.c.}
                        \right \}
\nonumber \\
         &+& \frac{U}{N_k}
             \bigl (
                     d^{\dagger} d - 1/2
             \bigr )
             \sum_{k, k'}
                   :\! c^{\dagger}_{k} c^{}_{k'}\!\!:\ ,
\end{eqnarray}
where $N_k$ is the number of $k$ points that are summed
over. This form of the Hamiltonian differs from that
of Eq.~(\ref{H-eff}) in the normalization factors
multiplying $t$ and $U$. In Eq.~(\ref{H-eff}), the
summations over the momenta are restricted to
$N' < N$ values of $k$, hence the conversion between
$(t_{\rm eff}, U_{\rm eff})$ and $(t, U)$ reads as
\begin{equation}
t = t_{\rm eff} \sqrt{ \frac{N'}{N} }
\;\; , \;\;\;\;\;
U = U_{\rm eff} \frac{N'}{N} .
\end{equation}
For the linear dispersion considered here, the
ratio $N'/N$ equals $D_{\rm eff}/D$, where
$D_{\rm eff} \sim \omega_0$ is the effective
bandwidth in the Hamiltonian of Eq.~(\ref{H-eff})
and $D$ is the full bandwidth pertaining to the
original electron-phonon Hamiltonian of
Eq.~(\ref{H-ep}). Thus, the conversion between
the corresponding coupling constants becomes
\begin{equation}
t = t_{\rm eff} \sqrt{ \frac{D_{\rm eff}}{D} }
\;\; , \;\;\;\;\;
U = U_{\rm eff} \frac{D_{\rm eff}}{D} ,
\label{conversion-1}
\end{equation}
which is naturally accounted for, as we shall
see, in the NRG.

\subsection{Finite-size spectrum of the IRLM near the
            free-impurity fixed point}

\subsubsection{Fixed-point spectrum for $t = U = 0$}

Consider first the noninteracting free-impurity fixed
point of the IRLM with $t = U = 0$. Since the impurity
level is decoupled from the band, the fixed-point
spectrum is simply that of a free symmetric band,
with an extra degeneracy of two due to the impurity
level which can be either empty or full. Thus, the
ground state is doubly degenerate for odd NRG
iterations $N$, and four-fold degenerate for even
iterations. The corresponding eigenstates are
conveniently labeled by a pair of numbers
\begin{equation}
Q_d = 2 d^{\dagger} d - 1
\end{equation}
and
\begin{equation}
Q_c = \sum_{n = 0}^{N}
            \bigl [
                    2 f^{\dagger}_{n} f^{}_{n} - 1
            \bigr ] ,
\end{equation}
which serve as good quantum numbers for zero tunneling
(only their sum $Q_d + Q_c$ is conserved for nonzero
$t$). The two degenerate ground states for
odd $N$ correspond to $(Q_d, Q_c) = (\pm 1, 0)$,
while the four degenerate ground states for even $N$
are labeled by $(Q_d, Q_c) = (\pm 1, \pm 1)$. The
extra two-fold degeneracy for even $N$ stems from
the presence of a conduction-electron mode that lies
exactly at the Fermi level. The fixed-point spectra
for odd and even iterations are listed in
Tables~\ref{Table-odd} and \ref{Table-even} up to
the second excitation energy.

\begin{table}
\begin{tabular}{c|c|c}
        \hline
        Energy level  & Quantum numbers
                      & Hopping matrix \\
        ($\Lambda = 2$) & $(Q_d ,Q_c)$ & element to \\
        \hline
        \multirow{4}{0.2cm}[-0.1cm]{0}
                                & $(-1,-1)$ & --- \\
                                & $(+1,-1)$ & $(-1,+1)$ \\
                                & $(-1,+1)$ & $(+1,-1)$ \\
                                & $(+1,+1)$ & --- \\
        \hline
        \multirow{8}{1cm}[-0.2cm]{0.9723}
                                & $(-1,-3)$ & --- \\
                                & $(+1,-3)$ & $(-1,-1)$ \\
                                & $(-1,-1)$ & $(+1,-3)$ \\
                                & $(+1,-1)$ & --- \\
                                & $(-1,+1)$ & --- \\
                                & $(+1,+1)$ & $(-1,+3)$ \\
                                & $(-1,+3)$ & $(+1,+1)$ \\
                                & $(+1,+3)$ & --- \\
        \hline
        \multirow{4}{1cm}[-0.1cm]{1.9446}
                                & $(-1,-1)$ & --- \\
                                & $(+1,-1)$ & $(-1,+1)$ \\
                                & $(-1,+1)$ & $(+1,-1)$ \\
                                & $(+1,+1)$ & --- \\ \hline
\end{tabular}
\caption{Same as Table~\ref{Table-odd} for even
         iteration number $N$.
        }
\label{Table-even}
\end{table}

\subsubsection{Fixed-point spectrum for $t = 0$
               and $U \neq 0$}

\begin{figure}[tb]
\centerline{
\includegraphics[width=85mm]{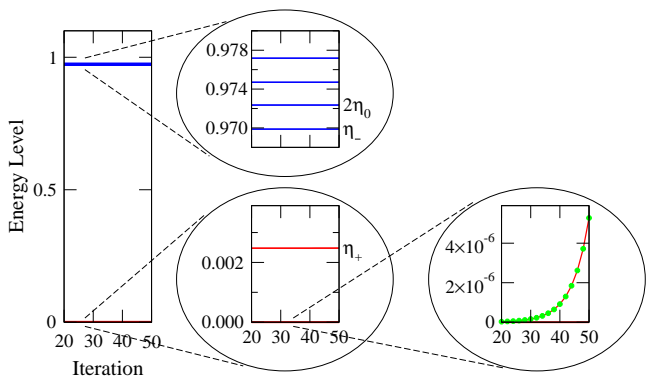}
}\vspace{0pt}
\caption{(Color online)
         Finite-size spectrum of the IRLM near the
         free-impurity fixed point, for $U/D = 0.01$,
         $t/D = 10^{-9}$, $\Lambda = 2$, and even
         iterations $N$. A nonzero $U$ lifts the
         four-fold and eight-fold degeneracies of
         the ground state and the first excitation
         energy, respectively, splitting them into
         distinct doublets (middle two panels).
         An infinitesimal $t$ further splits some
         of the doublets. In particular, the
         ground-state doublet is split according
         to $\gamma (t/D) \Lambda^{N (1 + \alpha)/4}$
         (right panel, green dots), where
         $\gamma$ equals $0.835$ and $\alpha$ is
         given by Eq.~(\ref{alpha}). The elementary
         particle, hole, and particle-hole excitations
         in the $Q_{d} = 1$ sector are marked by
         $\eta_{+}$, $\eta_{-}$, and $2\eta_0$ in the
         middle panels (see text for details).}
\label{Fig:fig_IRLM}
\end{figure}

A small but finite $U$ lifts certain degeneracies of
the $U = 0$ spectrum while maintaining the same pair
of quantum numbers $(Q_d, Q_c)$. In particular, the
ground-state quartet for even $N$ is split into two
doublets, each composed of two particle-hole-symmetric
states. The order of doublets discloses the sign of
$U$. When $U > 0$, the states labeled by $(+1, -1)$
and $(-1, +1)$ [$(+1, +1)$ and $(-1, -1)$] form the
ground-state [excited] doublet, while the order is
reversed for $U < 0$. The magnitude of $U$ can be
deduced in turn from a standard phase-shift
analysis~\cite{ALPC92} (see below), which requires
identification of the elementary particle, hole,
and particle-hole excitations. For $U > 0$ and even
$N$, these are given by the excitation energies
$\eta_{+}$, $\eta_{-}$, and $2\eta_0$ of the
$Q_d = 1$ sector, depicted in the middle panels
of Fig.~\ref{Fig:fig_IRLM}.
In the $Q_d = -1$ sector the elementary particle
and hole excitations are interchanged, reflecting
a reversal in sign of the scattering potential
experienced by the band electrons when the level
is empty.

\subsubsection{Spectrum in vicinity of the
               free-impurity fixed point}

In contrast to $U$, which is a marginal perturbation,
the tunneling term $t$ is a relevant one, driving
the system away from the free-impurity fixed point
to a new strong-coupling fixed point with $\pi/2$
phase shift of the scattered electrons. The new
fixed-point spectrum differs substantially from the
$t = 0$ one. However, it takes these differences
a while to develop in the course of the NRG
iterations. Starting with an infinitesimal tunneling
amplitude, the finite-size spectrum displays only
miniscule deviations from the $t = 0$ one over
many iterations. This behavior persists as long as
the renormalized tunneling amplitude remains small.
We focus hereafter on $U > 0$ (the case relevant
to phonon-assisted tunneling) and on this portion
of the NRG level flow.

The main effect of $t$ at these iterations is to
lift some of the remaining degeneracies of the
$t = 0$ fixed-point spectrum. For even $N$, certain
levels are split linearly in $t$, most notably the
ground-state doublet. The splittings for odd
$N$ are of higher order in $t$, reflecting the
absence of a direct tunneling matrix element between
degenerate eigenstates of the $t = 0$ spectrum. (A
detailed analysis of these matrix elements is
presented in the right-hand-side columns of
Tables~\ref{Table-odd} and \ref{Table-even}.) As
we next describe, splitting of the ground-state
doublet for even $N$ is proportional to
$\tilde{t}(D_N) = t(D_N)/D_N$, where $t(D_N)$ is
the renormalized tunneling amplitude at energy $D_N$.

According to a perturbative renormalization-group
(RG) analysis of the IRLM,~\cite{BVZ07}
the dimensionless tunneling amplitude
$\tilde{t} = t/D$ obeys the RG equation
\begin{equation}
\frac{d \tilde{t}}{d l} = \frac{1}{2}
                   (1 + \alpha) \tilde{t}
\label{IRLM-RG}
\end{equation}
with
\begin{equation}
\alpha = \frac{4}{\pi} \delta_{\rm U}
         - \left (
                   \frac{2}{\pi} \delta_{\rm U}
           \right )^2 .
\label{alpha}
\end{equation}
Here, $\delta_{\rm U} = \arctan (\pi \rho_0 U/2)$ is
the phase shift associated with $U$ in the absence of
tunneling, $\rho_0 = 1/(2 D)$ is the conduction-electron
density of states, and $l$ equals $\ln (D/D')$ with $D'$
the running bandwidth. Note that Eq.~(\ref{IRLM-RG})
is perturbative in $\tilde{t}$ but includes all
orders in $U$. It is supplemented in principle by a
second RG equation describing the renormalization
of $\alpha$,~\cite{BVZ07} yet the latter equation
can be neglected when $\tilde{t}$ is small.
Solution of Eq.~(\ref{IRLM-RG}) yields the
renormalized tunneling amplitude at energy $D'$:
\begin{equation}
\tilde{t}(D') = \tilde{t}(D)
                \left (
                        \frac{D}{D'}
                \right )^{(1 + \alpha)/2} .
\end{equation}
In the context of the NRG, $D/D'$ is replaced at
iteration $N$ with $D_0/D_N = \Lambda^{N/2}$,
resulting in
\begin{equation}
\tilde{t}(D_N) = \Lambda^{N (1 + \alpha)/4}
                 \left (
                         \frac{t}{D}
                 \right ) .
\label{t_D_N}
\end{equation}

We have found empirically that the ground-state
splitting for even $N$ is proportional to
$\tilde{t}(D_N)$, provided the latter coupling is not
too large. Here, by ground-state splitting we refer to
the difference in energy between the ground state and
the first excited state of the $Q = Q_{d} + Q_{c} = 0$
sector [the latter state is no longer the lowest
global excitation~\cite{Comment-on-levels} if
$\tilde{t}(D_N) > U/D_N$]. The accuracy of our
statement is demonstrated in the right-most panel
of Fig.~\ref{Fig:fig_IRLM}, where the ground-state
splitting (red line) is compared for $U/D = 0.01$
to $\gamma \tilde{t}(D_N)$ with $\gamma = 0.835$
(green dots). Excellent agreement is obtained. A
similar degree of accuracy extends to all values
of $U > 0$, provided $\tilde{t}(D_N) < 0.1$. For
larger values of $\tilde{t}(D_N)$, the linear relation
between $\tilde{t}(D_N)$ and the ground-state splitting
gradually breaks down due to the departure from
weak coupling. As for $\gamma$, its value depends
on both $U$ and $\Lambda$. For $\Lambda = 2$, the
discretization parameter used throughout this
work, $\gamma$ grows monotonically from $0.835$ to
$0.94$ in going from $U = 0$ to $U \to \infty$.
For $0 < U/D < 0.3$, the regime of relevance to our
discussion, $\gamma$ changes by no more than 4\%,
allowing one to use the single figure $\gamma = 0.835$
in order to extract $\tilde{t}_{\rm eff}$.

\subsection{Extracting the couplings $t_{\rm eff}$
            and $U_{\rm eff}$}

Our analysis thus far has provided us with a thorough
understanding of the finite-size spectrum of the IRLM
near the free-impurity fixed point. Next we specify
how one can exploit this knowledge to extract the
coupling constants $t_{\rm eff}$ and $U_{\rm eff}$
that enter the effective IRLM of Eq.~(\ref{H-eff}).

As anticipated in Sec.~\ref{Sec:Extended-AAL}, the
finite-size spectrum of the Hamiltonian of
Eq.~(\ref{H-ep}) is found to be well described
for $\Gamma_{\rm eff} \ll D_N \ll \omega_0$ by
the weak-coupling spectrum of the IRLM with $U > 0$.
The sign of $U$ is exposed from the nondegenerate
ground state for even iterations $N$, which
belongs to the $Q = 0$ sector.~\cite{Comment-on-U}
To extract the model parameters of the effective
IRLM Hamiltonian we have implemented the following
procedure. First, a particular iteration number
$N = 22$ was chosen such that $D_N = 0.146 \omega_0$.
The dimensionless tunneling amplitude
$\tilde{t}(D_N) = \Delta E/\gamma$ was
next extracted from the energy splitting $\Delta E$
between the ground state and the first excited
state of the $Q = 0$ sector. The value of $\gamma$
was set equal to $\gamma = 0.835$, in accordance with
the discussion above. Lastly, the dimensionless
Coulomb repulsion $\tilde{U}(D_N) = U(D_N)/D_N$
was extracted from the elementary particle, hole, and
particle-hole excitations $\eta_{+}$, $\eta_{-}$, and
$2\eta_0$ according to the standard
prescription~\cite{ALPC92}
\begin{equation}
\tilde{U} = \frac{4}{\pi} \tan (\delta) ,
\label{U-from-delta}
\end{equation}
with
\begin{equation}
\delta =
\frac{\eta_{-} - \eta_{+}}{2 \eta_0} + \frac{\pi}{2} .
\label{delta-from-eta}
\end{equation}

The procedure outlined above provided us with
estimates for $\tilde{t}$ and $\tilde{U}$ at the
energy scale $D_{\rm eff} = D_{N = 22}$. The accuracy
of the couplings so obtained can be summarized
as follows. If less than $0.1$, the dimensionless
tunneling amplitude $\tilde{t}$ is accurate
to within about 4\%, provided $\tilde{U}$ is
simultaneously smaller than $0.3$. This
criterion was met for all points displayed
in Fig.~\ref{Fig:t_eff}. Accuracy of the
dimensionless Coulomb repulsion $\tilde{U}$
is controlled in turn by the ratio
$\tilde{t}/\tilde{U}$. Indeed,
Eqs.~(\ref{U-from-delta}) and (\ref{delta-from-eta})
are precise for $\tilde{t} = 0$, acquire
a small correction proportional to
$\tilde{t}/\tilde{U}$ when $\tilde{t}$ is
small, and break down as soon as $\tilde{t}$
approaches $\tilde{U}$. All points displayed
in Fig.~\ref{Fig:U_eff} fall in the range
$\tilde{t}/\tilde{U} < 2 \times 10^{-3}$,
corresponding to an accuracy of order 1\%
for $\tilde{U}$.

Finally, the conversion from the dimensionless
couplings $\tilde{t}$ and $\tilde{U}$ to
$t_{\rm eff}$ and $U_{\rm eff}$ proceeds as follows.
Using the notation of Eq.~(\ref{conversion-1}),
one has that
\begin{equation}
t_{\rm eff} = t \sqrt{ \frac{D}{D_{\rm eff}} }
            = D_{\rm eff} \tilde{t}
              \sqrt{ \frac{D}{D_{\rm eff}} }
\end{equation}
and
\begin{equation}
U_{\rm eff} = \frac{D}{D_{\rm eff}} U
            = D \tilde{U} .
\end{equation}
Since $D_{\rm eff}/D$ is replaced at the $N$th
iteration of the NRG with $D_{N}/D_0 = \Lambda^{-N/2}$,
we arrive at
\begin{equation}
\frac{ t_{\rm eff} }{D} = \tilde{t} \Lambda^{-N/4}
\;\; , \;\;\;\;\;
\frac{ U_{\rm eff} }{D} = \tilde{U} .
\end{equation}


\begin{thebibliography}{9}
\bibitem{comment-on-shift-I}
         The transformation also generates the constant
         term $\epsilon_d^{} -\lambda^2 \omega_0 (1 - 2N_0)
         = \epsilon_d^{} - \epsilon_d^{\ast}$, which
         uniformly shifts the spectrum of the Hamiltonian.
\bibitem{comment-on-DOS}
         We use the convention $\rho(\epsilon) = (1/N)
         \sum_k \delta(\epsilon - \epsilon_k)$ for the
         conduction-electron density of states, such
         that $\rho(\epsilon)$ is normalized to unity.
         In case of a symmetric rectangular density of
         states, one therefore has that $\rho_0 = 1/(2 D)$.
\bibitem{comment-on-resonance}
         Note that the exact resonance condition is
         $\epsilon^{}_d = \epsilon_d^{\ast} =
         \lambda^2 \omega_0 (1 - 2N_0)$, which differs
         from $\epsilon^{}_d = 0$ when $N_0 \neq 1/2$.
         Nevertheless, this distinction enters
         $\sigma(\omega + i\eta)$ only at higher orders
         in $\lambda$.
\bibitem{Comment-on-U}
         For $U < 0$ the ground state is a doublet,
         consisting of two particle-hole-symmetric
         states with $Q = \pm 2$.
\end{thebibliography}
\end{document}